\documentclass[journal=jctcce,manuscript=article,layout=onecolumn]{achemso}
\setkeys{acs}{articletitle = true, chaptertitle = true}

\usepackage[version=3]{mhchem} 
\usepackage[T1]{fontenc}       
\usepackage{mathptmx}

\makeatletter
\let\l@addto@macro\relax
\makeatother
\usepackage[fontsize=12pt]{scrextend}

\usepackage{graphicx}  
\usepackage{bm}        
\usepackage{bbm}       
\usepackage{amssymb}   
\usepackage{amsmath}   
\usepackage{physics}   
\usepackage{textgreek} 
\usepackage{subfigure} 
\usepackage{multirow}  
\usepackage{booktabs}  
\usepackage{multicol}
\usepackage{tabularx}
\usepackage{color, colortbl} 
\definecolor{Gray}{gray}{0.9} 
\definecolor{LightGreen}{rgb}{0.96, 0.99, 0.96} 
\usepackage{threeparttable} 
\usepackage{hyperref}  
\hypersetup{colorlinks=true, linkcolor=black, urlcolor=black, citecolor=black, plainpages=false,
bookmarksnumbered=true}
\usepackage{xurl} 
\usepackage{longtable} 
\usepackage{rotating}  
\usepackage{siunitx}   
\usepackage{xcolor}     
\usepackage{cases}     
\usepackage[utf8]{inputenc} 
\usepackage{adjustbox}

\newcommand{\RN}[1]{\textup{\uppercase\expandafter{\romannumeral#1}}}

\author{Manas Sharma}
\email{manassharma@iisc.ac.in}
\affiliation{Department of Chemical Engineering, Indian Institute of Science, Bengaluru, Karnataka 560012, India}

\author{Sudeep Punnathanam}
\affiliation{Department of Chemical Engineering, Indian Institute of Science, Bengaluru, Karnataka 560012, India}

\author{Ananth Govind Rajan}
\email{ananthgr@iisc.ac.in}
\affiliation{Department of Chemical Engineering, Indian Institute of Science, Bengaluru, Karnataka 560012, India}

\title{MLIP Studio: An Open Platform for Interactive Benchmarking and Atomistic Simulations Using Machine Learning Interatomic Potentials}

\begin{document}

\setlength{\fboxrule}{0pt}
\begin{tocentry}
    \includegraphics[width=10cm]{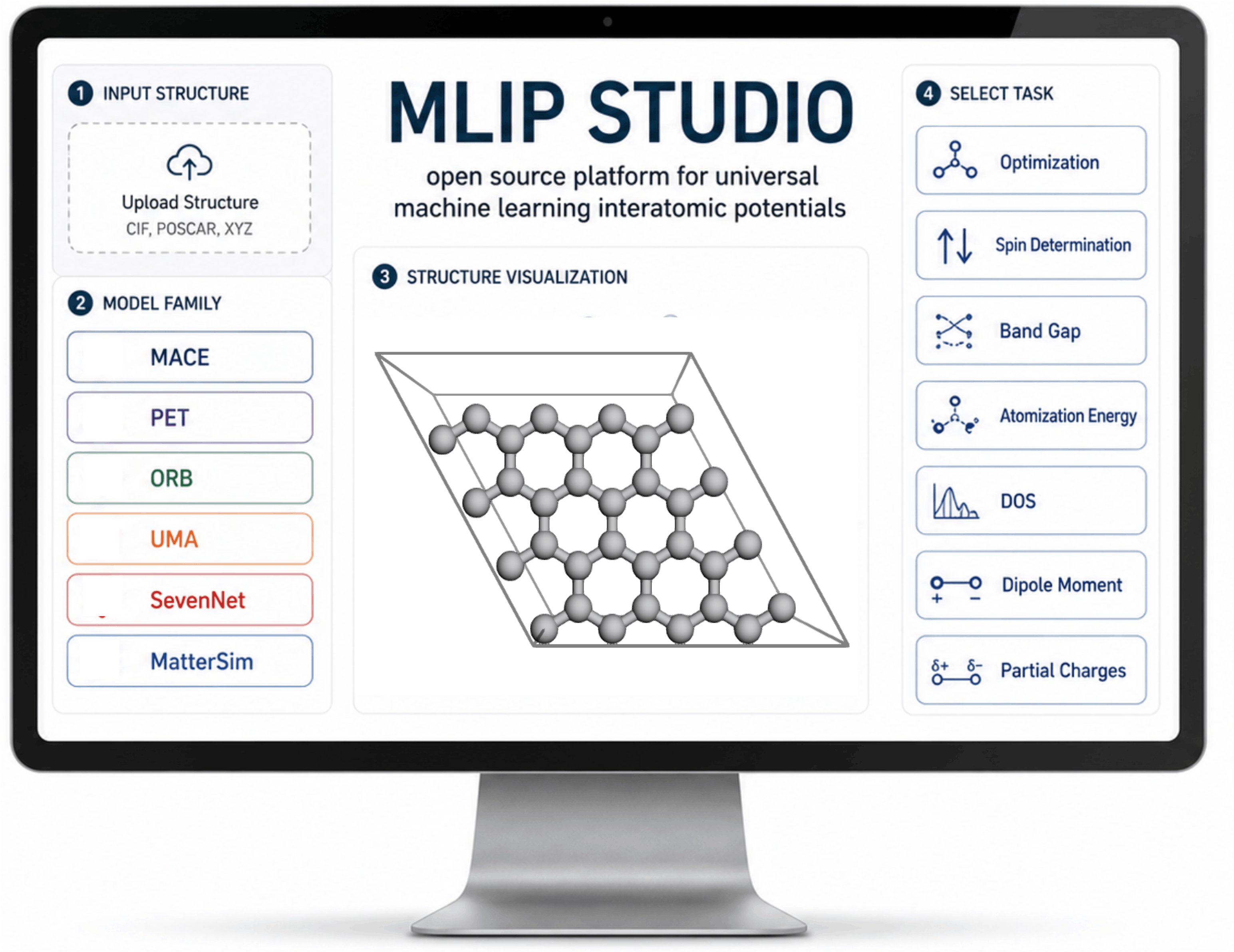}
\end{tocentry}

\begin{abstract}
    Universal machine learning interatomic potentials (MLIPs) are foundation AI models transforming atomistic simulations, but their practical use remains hindered by fragmented software ecosystems, dependency conflicts, and the lack of accessible benchmarking tools. These models approach first-principles density functional theory (DFT) accuracy at a fraction of the computational cost. 
    We introduce MLIP Studio (available at \url{https://mlipstudio.iisc.ac.in}), an open and free platform that brings more than 60 universal MLIPs into a unified interactive interface for molecules and materials. The platform enables end-to-end MLIP-driven workflows, including property prediction, geometry optimization, vibrational and equation-of-state analysis, spin-state determination, custom model deployment, and high-throughput benchmarking against reference data. Automated parity plots and sortable error tables facilitate rapid identification of element-wise outliers and problematic data points. We demonstrate that MLIP-based pre-optimization can reduce subsequent DFT optimization effort by $\sim 33\times$. Additionally, the application enables benchmarking of computational performance. Through a comprehensive case study involving the 2D magnetic material CrCl$_3$ on a sapphire substrate, we show how cross-model comparisons of various properties and potential-energy landscapes can guide task-specific MLIP selection. Overall, MLIP Studio lowers the barrier to the reliable use of foundation models in end-to-end research workflows, benchmarking, and education in computational chemistry and materials science.
\end{abstract}

\section{Introduction}
\label{sec:introduction}
Machine learning interatomic potentials (MLIPs) are data-driven models that approximate the quantum-mechanical potential energy surface~\cite{jacobs2025practical, wan2024construction, wang2024machine}. 
In contrast to empirical force fields, which rely on fixed functional forms with limited flexibility, MLIPs employ flexible neural network architectures that can systematically approach the accuracy of the target quantum-mechanical method on which the model is trained.
Quantum-mechanical density functional theory (DFT) calculations are the workhorse of modern-day computational materials science~\cite{KSDFT_1965, verma2020status, bursch2022best, norskov2011density, teale2022dft}. Though highly accurate, the computationally expensive nature of DFT~\cite{DFT_test_1996, DFT_test_2004, DFT_test_2006, DFT_test_2009, DFT_perspective_2012} has motivated the development of MLIPs that combine near ab‑initio accuracy with orders‑of‑magnitude improvements in computational speed.
Over recent years, a variety of universal or foundation MLIPs have emerged that can be applied across a broad range of chemical systems, from simple molecules to complex solids~\cite{yuan2026foundation,rhodes2025orb, neumann2024orb, wood2025family, yang2024mattersim, fu2025learning, batatia2022mace, park2024scalable, batzner20223}. 
Universal MLIPs are large-scale interatomic models pretrained on chemically diverse DFT datasets with millions of datapoints to achieve transferable, zero-shot accuracy across molecules, solids, and interfaces without system-specific retraining.
The burgeoning field of universal MLIPs is rapidly transforming atomistic simulations in chemistry and materials science~\cite{yuan2026foundation}. 

Despite rapid architectural innovation, the MLIP ecosystem remains fragmented across independent software stacks, incompatible library dependencies, and heterogeneous interfaces. 
As a result, systematic benchmarking and reproducible comparison of models developed by different groups remain technically challenging. 
Furthermore, while there are several interactive applications for quantum-mechanical or classical molecular calculations (e.g., TMoleX~\cite{steffen2010tmolex}, GaussView~\cite{gaussview6}, BURAI~\cite{burai}, Quantum ATK~\cite{smidstrup2020quantumatk}, and Avogadro~\cite{hanwell2012avogadro}), comparable platforms for MLIPs remain largely absent, limiting the accessibility of universal MLIPs for educational training and rapid prototyping.
MLIP Studio addresses these challenges by providing a free, source-available, and fully interactive interface that permits users to upload or import any chemical structure, select from among more than sixty universal MLIP models, and perform various calculations, including energy, force, stress, bandgap, and dipole moment prediction, geometry optimization, vibrational analysis, and the evaluation of bulk modulus and cohesive and atomization energies, all within a single, seamlessly integrated environment.

The foundation of MLIP Studio lies in the philosophy of accessible scientific software, a philosophy that is gaining momentum in the broader materials informatics community~\cite{lehtola2022free,jain2020materials, horton2025accelerated}. 
Several recent initiatives, such as MLIP Arena~\cite{chiang2025mlip} and MatBench Discovery~\cite{riebesell2023matbench}, have demonstrated the importance of fairness and transparency in MLIP benchmarking by considering evaluations that incorporate physical consistency, transferability, and real‑world applicability.
At the same time, a host of interactive web applications, built with modern frameworks such as Streamlit~\cite{khorasani2022web}, have emerged across domains ranging from drug discovery to materials property prediction. 
For example, Materials Project's \cite{jain2020materials, jain2013commentary, horton2025accelerated} web-based application provides a searchable database of crystal structures and their computed thermodynamic and electronic properties, along with APIs and visualization tools. 
Similarly, C2DB \cite{gjerding2021recent} provides a web-based platform to explore and visualize data for two-dimensional materials, while Materials Cloud \cite{talirz2020materials} offers curated datasets, interactive structure viewers, and reproducible computational workflows. 
Together, these web-based platforms make materials data and related tools easily accessible for exploration and analysis.
Despite these significant developments, there remain very few freely accessible platforms that provide comprehensive MLIP-oriented workflows while supporting models developed by multiple research groups. 
Most existing solutions are proprietary commercial services, such as RowanSci \cite{rowan_scientific_platform_2026}, Quantistry\cite{quantistry_materials_intelligence_2026}, Matalantis \cite{matlantis_atomistic_simulator_2026}, and QpiAI\cite{qpiai_tech_2026}, which rely on closed-source implementations and have high subscription costs.

MLIP Studio is designed to fill this gap by leveraging freely available tools to provide a universal and extendable web‑based framework for MLIP evaluation and MLIP-based atomistic simulations. 
MLIP Studio provides access to sixty-two universal MLIPs from five major model families: multi-atomic cluster expansion (MACE)~\cite{batatia2022mace}, FairChem~\cite{wood2025family}, MatterSim~\cite{yang2024mattersim}, orbital materials (ORB)~\cite{neumann2024orb,rhodes2025orb}, SevenNet~\cite{park2024scalable}, and point-edge transformer (PET)~\cite{pozdnyakov2023smooth,mazitov2025pet}, within the same computational environment. 
The web application is built using Streamlit~\cite{khorasani2022web}, a Python‑based framework known for its rapid prototyping capabilities and ease of integration with powerful back‑end libraries such as the Atomic Simulation Environment (ASE)~\cite{hjorth2017atomic} and Py3Dmol~\cite{seshadri20203dmol}, which allows for interactive three‑dimensional visualization of molecular structures. 
This easy-to-use graphical user interface (GUI) ensures that MLIP-powered atomistic simulations are accessible to researchers with minimal coding expertise.
The use of Python as the backend language ensures that the underlying architecture is modular and easily extendable, and works well with the existing ecosystem of diverse datasets, materials informatics libraries, and MLIP frameworks.
As a result, users can seamlessly switch between MLIP models and directly compare their performance on their own datasets, without writing any code.

Unlike commercial alternatives, which impose licensing fees and offer a limited choice of MLIPs, MLIP Studio is provided entirely free under the Academic Software License (ASL). 
The full source code is available on our GitHub~\cite{mlip_studio_github_2025} repository.
Additionally, we host a dedicated instance at \href{https://mlipstudio.iisc.ac.in}{https://mlipstudio.iisc.ac.in}, running on our own server (with GPU acceleration), to provide a reliable and persistent access point for the community for demonstration purposes (with limits on system sizes). 
However, users can clone the repository and run it on their local machine to circumvent the limits placed on the free online version.
This commitment to openness fosters reproducibility, community-driven development, and transparent benchmarking across the growing ecosystem of universal MLIPs.

Beyond licensing and accessibility concerns, a significant practical barrier to MLIP adoption is the difficulty of installing and managing multiple MLIP frameworks within a single computational environment. 
Popular MLIP packages from FAIRChem~\cite{wood2025family,fu2025learning}, MACE\cite{batatia2022mace}, and ORB\cite{rhodes2025orb, neumann2024orb}, often depend on specific and mutually incompatible versions of core libraries like PyTorch, PyTorch Geometric, NumPy, or e3nn~\cite{geiger2022e3nn}. 
In practice, this means that a researcher wishing to compare predictions across several MLIPs must maintain separate virtual environments for each model, activating and switching between them for every new calculation, which is tedious and hampers the setup of a seamless workflow for systematic benchmarking. 
MLIP Studio eliminates this friction by providing a carefully curated environment in which all supported MLIPs are installed with mutually compatible library versions. 
This was achieved through extensive compatibility testing, in which we systematically identified library versions that, while not always matching those specified in each package's default requirements file, nonetheless produce correct and validated results across all supported models. 
Users can therefore switch between over 60 models instantaneously within a single session, without concern for dependency conflicts or environment management, enabling seamless and fair side‑by‑side comparisons that would otherwise require substantial setup effort.

The platform's modular, plugin-based architecture is designed to accommodate the rapid pace of MLIP development. 
New models, simulation protocols, and post‑processing tools can be integrated with minimal effort, ensuring that MLIP Studio remains current as new methods and datasets such as MAD~\cite{malosso2026high, mazitov2025massive, how2026universal}, OMAT24~\cite{barroso2024open}, OMOL25~\cite{levine2025open}, MaPES \cite{kaplan2025foundational}, and MPTraj~\cite{jain2020materials, jain2013commentary, horton2025accelerated}, continue to emerge. 
By supporting over sixty MLIP models spanning diverse architectural families in one place, the platform enables side‑by‑side comparisons that go beyond standard error metrics to include assessments of physical consistency, stability, and transferability based on users' own datasets.

Furthermore, user experience and accessibility have been key design criteria in the development of MLIP Studio. 
Recognizing that robust computational tools are only as effective as their usability, the web app has been engineered to support a minimal learning curve for end users. 
The integration of a drag‑and‑drop interface for file uploads, combined with intuitive selection menus for choosing simulation protocols and MLIP models, ensures that even users with limited computational backgrounds can harness the power of universal MLIPs to explore complex chemical phenomena. 
Therefore, beyond research workflows, MLIP Studio also serves as a pedagogical platform. 
By providing real-time visualization of geometry optimization trajectories, vibrational spectra, equation-of-state fitting, and spin-state scans, the application allows users to interactively explore core atomistic simulation concepts without writing code. 
This makes the platform suitable for classroom demonstrations and graduate-level computational chemistry and materials science courses.

In addition to improving accessibility and unifying diverse MLIP frameworks, MLIP Studio is designed with a strong emphasis on practical workflows and model evaluation. 
It facilitates MLIP-based geometry pre-optimization, which yields structures close to the DFT minimum, significantly reducing subsequent first-principles optimization cost. 
The platform also enables rapid screening of configuration ensembles, where energy profiling and downloadable data allow straightforward identification of lowest-energy structures for further simulations. 
When reference data are available, automatically generated parity plots and sortable error tables facilitate identification of high-error configurations and efficient diagnosis of problematic training frames. 
All calculations report wall times, enabling systematic benchmarking of model performance across hardware. 
The platform also supports spin-state determination, providing useful initial guesses for DFT calculations. 
Additionally, it facilitates cross-model comparisons for potential energy surfaces generated through different MLIPs.

The remainder of this manuscript is organized as follows. 
Section 2 introduces the conceptual foundations of universal MLIPs and briefly summarizes the architectural families supported in the platform. 
Section 3 describes the general workflow of the web application, including structure input, model selection, and task configuration. 
Section 4 details the supported simulation tasks and illustrates representative applications. 
Additionally, Section 4 presents computational performance benchmarks on CPU and GPU hardware. 
Section 5 demonstrates an end-to-end case study involving the precursor for the 2D magnetic material CrCl$_3$ on a sapphire substrate, which combines these capabilities for model evaluation and selection on a realistic molecule--surface system.
Finally, Section 6 concludes with an outlook on future developments and extensions.

\section{Universal Machine Learning Interatomic Potentials Supported by MLIP Studio}

An MLIP learns a mapping from atomic numbers and Cartesian coordinates to quantum-mechanical observables of interest.
Usually, the primary target is the potential energy surface. 
In conservative formulations, forces are obtained as the negative gradient of the predicted potential energy; in direct-force formulations, forces are predicted directly (without differentiation) for improved throughput and reduced memory footprint.
Beyond energies and forces, MLIPs can also be trained to predict properties such as dipole moments, band gaps, partial charges, density of states (DOS), and so on~\cite{how2026universal}.
Most MLIPs developed by research groups for a particular problem are bespoke: they are trained for a specific material, phase, or reaction pathway and optimized for accuracy within that restricted chemical space. 
Their predictive reliability typically degrades under distribution shifts, requiring retraining or active learning when new environments are encountered.
Universal or foundation MLIPs pursue a different objective. 
They are pretrained on chemically diverse datasets comprising millions of quantum-mechanical DFT calculations. 
The goal is transferable accuracy across broad regions of the chemical space without system-specific retraining. 
In this sense, a universal MLIP functions as a pretrained foundation model that can be deployed in a zero-shot manner or finetuned for downstream tasks.

Architecturally, most modern MLIPs are geometric graph neural networks (GNNs)~\cite{yuan2026foundation}. 
In this formulation, an atomic configuration is represented as a graph whose nodes correspond to atoms and whose edges connect neighboring atoms within a finite radial cutoff. 
Message passing layers iteratively update atomic feature vectors by aggregating information from neighboring atoms, thereby constructing many-body representations of the local chemical environment. 
The radial cutoff defines the spatial extent of interactions included in the model and ensures linear scaling with system size by restricting computation to local neighborhoods~\cite{duval2023hitchhiker}.

Physical symmetries impose additional constraints on the predictions made by MLIPs. 
The predicted total potential energy must be invariant to global translations, rotations, and permutations of atoms, while forces must transform equivariantly under rotations~\cite{duval2023hitchhiker}. 
State-of-the-art MLIPs enforce these symmetries either explicitly through equivariant tensor representations~\cite{batzner20223,batatia2022mace} using spherical harmonics
and Clebsch-Gordan coefficients to represent and update atomic environments (e.g., MACE~\cite{batatia2022mace} and NequIP~\cite{geiger2022e3nn}) or implicitly through data augmentation and regularization (equigrad)~\cite{rhodes2025orb}.
For a systematic explanation of geometric GNN architectures, symmetry handling, tensor representations, and their mathematical foundations, we refer the reader to the detailed review in ref. \citenum{duval2023hitchhiker}.

Within this landscape, several architectural families of universal MLIPs have emerged that differ in how they enforce symmetry, scale model capacity, and optimize computational efficiency. 
Below, we briefly summarize the major model families supported by MLIP Studio and highlight the design choices that distinguish them.

\begin{itemize} 

\item \textbf{MACE:} MACE introduced higher-order equivariant message passing in which many-body correlations can be captured within each interaction block, building upon the atomic cluster expansion (ACE) formulation~\cite{drautz2019atomic}. 
By increasing body order without increasing network depth, MACE achieves high expressivity with only a few layers, improving data efficiency and parallel scalability. 
Its use of irreducible spherical tensor features ensures strict $E(3)$ equivariance while keeping parameter growth controlled relative to earlier tensor field networks~\cite{batatia2022mace}.

\item \textbf{ORB:} The ORBv3 family charts the performance-speed-memory Pareto frontier by offering both conservative models (that enforce rotational invariance through equigrad regularization) and lightweight direct-force models. 
The latter provides a $\approx10\times$ reduction in latency and $\approx8\times$ reduction in memory compared to traditional MLIPs, enabling stable molecular dynamics for systems exceeding 100,000 atoms~\cite{rhodes2025orb}.

\item \textbf{UMA (FAIRChem):} The universal model for atoms (UMA)~\cite{wood2025family} family is based on the equivariant smooth energy network (eSEN) architecture~\cite{fu2025learning} and leverages a mixture of linear experts (MoLE) architecture to scale model capacity (up to 1.4B parameters). 
Trained on a massive multi-domain dataset of 459 million atomic systems, UMA supports multi-task learning. 
Users can specify the target domain, such as \texttt{omat} (materials) \cite{barroso2024open}, \texttt{omol} (molecules) \cite{levine2025open}, \texttt{oc20} (catalysis), \texttt{omc} (molecular crystals), or \texttt{odac} (MOFs), via a task embedding that routes representations through specialized experts.
A UMA model with the \texttt{omol} task head can take in spin and charge as inputs in addition to the atoms and their positions.
\item \textbf{MatterSim (Microsoft):} Designed for generalization across extreme thermodynamic conditions, MatterSim \cite{yang2024mattersim} is trained on the \texttt{MPF-TP} dataset. 
This dataset covers temperatures from 0 to 5000 K and pressures up to 1000 GPa, and training is performed using an uncertainty-aware sampling and active learning framework.
This makes MatterSim particularly useful for various application areas, such as, lattice dynamics, mechanical properties, and thermodynamics.

\item \textbf{SevenNet:} This equivariant GNN potential is specifically optimized for large-scale molecular dynamics through a scalable parallel algorithm compatible with spatial decomposition. 
By restricting communication to the original cutoff radius while exchanging node features and energy gradients, SevenNet achieves over 80\% parallel efficiency in weak-scaling scenarios on multi-GPU clusters.

\item \textbf{PET:} 
PET~\cite{mazitov2025pet} adopts a fundamentally different design from equivariant tensor GNNs by using a transformer-based message passing architecture operating on directed edges. 
Instead of enforcing $E(3)$ symmetry explicitly through spherical tensor algebra (as in MACE or NequIP), PET is rotationally unconstrained and learns approximate invariance/equivariance through data augmentation. 
At each message-passing step, edge features corresponding to bonds are updated via attention-based transformer blocks, allowing flexible and effectively high-order interactions with a small number of layers. 
This yields a model that is a universal approximator even at shallow depth, with virtually unrestricted body order and angular resolution.
In practice, this leads to lower inference cost and favorable scaling, placing PET-based models on a competitive accuracy–efficiency Pareto frontier along with models such as MACE, SevenNet, and ORB. 
Additionally, PET models support both conservative energy-based training and direct-force prediction heads, similar in spirit to ORB, offering further flexibility in performance–accuracy trade-offs.
PET developers provide models trained on popular datasets such as MAD~\cite{mazitov2025massive,how2026universal,malosso2026high}, OMAT~\cite{barroso2024open}, and SPICE~\cite{eastman2023spice}.
\end{itemize}
\subsection{Available MLIP Models}

Table~\ref{tab1:models} provides a comprehensive list of the universal MLIPs supported by the platform, organized by their respective model families:

\begin{table*}[htbp]
\centering
\caption{Universal MLIP models supported in MLIP Studio, organized by family.}
\begin{adjustbox}{width=\textwidth}
\begin{tabular}{llllll}
\hline
\textbf{MACE}~\cite{batatia2022mace, batatia2026mace,kovacs2025mace, batatia2025foundation} & \textbf{FAIRChem (UMA/ESEN)}~\cite{wood2025family,fu2025learning} & \textbf{ORB v3}~\cite{rhodes2025orb} & \textbf{MatterSim}~\cite{yang2024mattersim} & \textbf{SevenNet}~\cite{park2024scalable,kim2024data,kim2026optimizing} & \textbf{PET}~\cite{mazitov2025pet,how2026universal,malosso2026high,pozdnyakov2023smooth} \\
\hline

MACE MPA Medium & UMA Small 1.1 & V3 OMOL Conservative & V1 SMALL & 7net-0 & PET-MAD-XS (v1; v1.5) \\
MACE OMAT Medium & UMA Small 1.2 & V3 OMOL Direct & V1 LARGE & 7net-l3i5 & PET-MAD-S (v1; v1.5) \\
MACE OMAT Small & ESEN MD Direct All OMOL & V3 OMAT Conservative (inf) &  & 7net-omat & PET-OAM-L \\
MACE MATPES r2SCAN Medium & ESEN SM Conserving All OMOL & V3 OMAT Conservative (20) &  & 7net-mf-ompa & PET-OMAT-XS \\
MACE MATPES PBE Medium & ESEN SM Direct All OMOL & V3 OMAT Direct (inf) &  &  & PET-OMAT-S \\
MACE MP 0a Small &  & V3 OMAT Direct (20) &  &  & PET-OMAT-M \\
MACE MP 0a Medium &  & V3 MPA Conservative (inf) &  &  & PET-OMAT-L \\
MACE MP 0a Large &  & V3 MPA Conservative (20) &  &  & PET-OMATPES-L \\
MACE MP 0b Small &  & V3 MPA Direct (inf) &  &  & PET-SPICE-S \\
MACE MP 0b Medium &  & V3 MPA Direct (20) &  &  & PET-SPICE-L \\
MACE MP 0b2 Small &  &  &  &  & PET-MAD-DOS \\
MACE MP 0b2 Medium &  &  &  &  & PET-OMAD-XS \\
MACE MP 0b2 Large &  &  &  &  & PET-OMAD-S \\
MACE MP 0b3 Medium &  &  &  &  & PET-OMAD-L \\
MACE ANI-CC Large (500k) &  &  &  &  &  \\
MACE OMOL-0 XL 4M &  &  &  &  &  \\
MACE OMOL-0 XL 1024 &  &  &  &  &  \\
MACE OFF 23 Large &  &  &  &  &  \\
MACE OFF 23 Medium &  &  &  &  &  \\
MACE OFF 23 Small &  &  &  &  &  \\
MACE OFF 24 Medium &  &  &  &  &  \\
MACE POLAR 1 Small &  &  &  &  &  \\
MACE POLAR 1 Medium &  &  &  &  &  \\
MACE POLAR 1 Large &  &  &  &  &  \\
\hline
\label{tab1:models}
\end{tabular}
\end{adjustbox}
\end{table*}

\section{General Workflow}
\label{sec:workflow}

The MLIP Studio web application is designed around a streamlined, four-step workflow: (i) structure input, (ii) model selection, (iii) task configuration, and (iv) execution and output visualization. 
Figure~\ref{fig:homepage} provides an annotated overview of the home page, highlighting each of these stages.

\subsection{Structure Input}
\label{sec:input}

The application supports multiple input methods, as shown in Figure~\ref{fig:homepage}a. 
Users may choose from the following options:

\begin{itemize}
    \item \textbf{Select Example:} A set of predefined example structures is provided, allowing users to quickly explore the application without preparing any input files.
    \item \textbf{Upload File:} Users can upload a structure file from their local device. 
    Supported file formats include CIF, extXYZ, XYZ, POSCAR, TURBOMOLE, and MOL.
    \item \textbf{Paste Content:} Instead of uploading a file, users may paste the contents of a structure file in any of the supported formats directly into a text box.
    \item \textbf{Materials Project ID:} A crystal structure can be imported directly from the Materials Project database~\cite{horton2025accelerated} by specifying its Materials Project identifier (mp-id).
    \item \textbf{PubChem:} Molecular structures can be imported from the PubChem database \cite{kim2016pubchem, kim2019pubchem, kim2023pubchem, kim2025pubchem} by providing either the molecule name or its chemical formula.
    \item \textbf{Batch Upload:} Multiple structure files in any combination of the supported formats can be uploaded simultaneously for batch processing.
    \item \textbf{extXYZ Trajectory Upload:} An extended XYZ trajectory file containing multiple frames or configurations can be uploaded for sequential evaluation.
\end{itemize}

Once a structure is provided through any of these methods, it is parsed and rendered in an interactive three-dimensional viewer using Py3Dmol~\cite{seshadri20203dmol}, as shown in Figure~\ref{fig:homepage}c. 
Users can rotate, zoom, and inspect the structure, and select from various visualization styles, such as, ball-and-stick, stick, or space-filling representations.

\subsection{Model Selection}
\label{sec:model_selection}

The model selection panel, shown in Figure~\ref{fig:homepage}b, allows users to choose from five major MLIP families (already described above) as well as semi-empirical (GFN2-xTB)~\cite{bannwarth2019gfn2} and classical (UFF)~\cite{rappe1992uff} calculators and a standalone D3 dispersion correction calculator~\cite{takamoto2021pfp}. 
Having a separate dispersion calculator is useful because not all models support a van der Waals dispersion correction option. Using this option, the dispersion correction can be applied later on for unsupported models as well.
Upon selecting a model family, the user is presented with the available models within that family. 
Additionally, for MACE-architecture models, users have the option to provide a custom model, either by uploading a model file or by specifying a URL from which the model can be downloaded. 
This enables the use of finetuned or in-development models within the same interface.

\subsection{Calculation Setup and Execution}
\label{sec:calc_setup}

\begin{figure}[h]
    \centering
    \includegraphics[width=\textwidth]{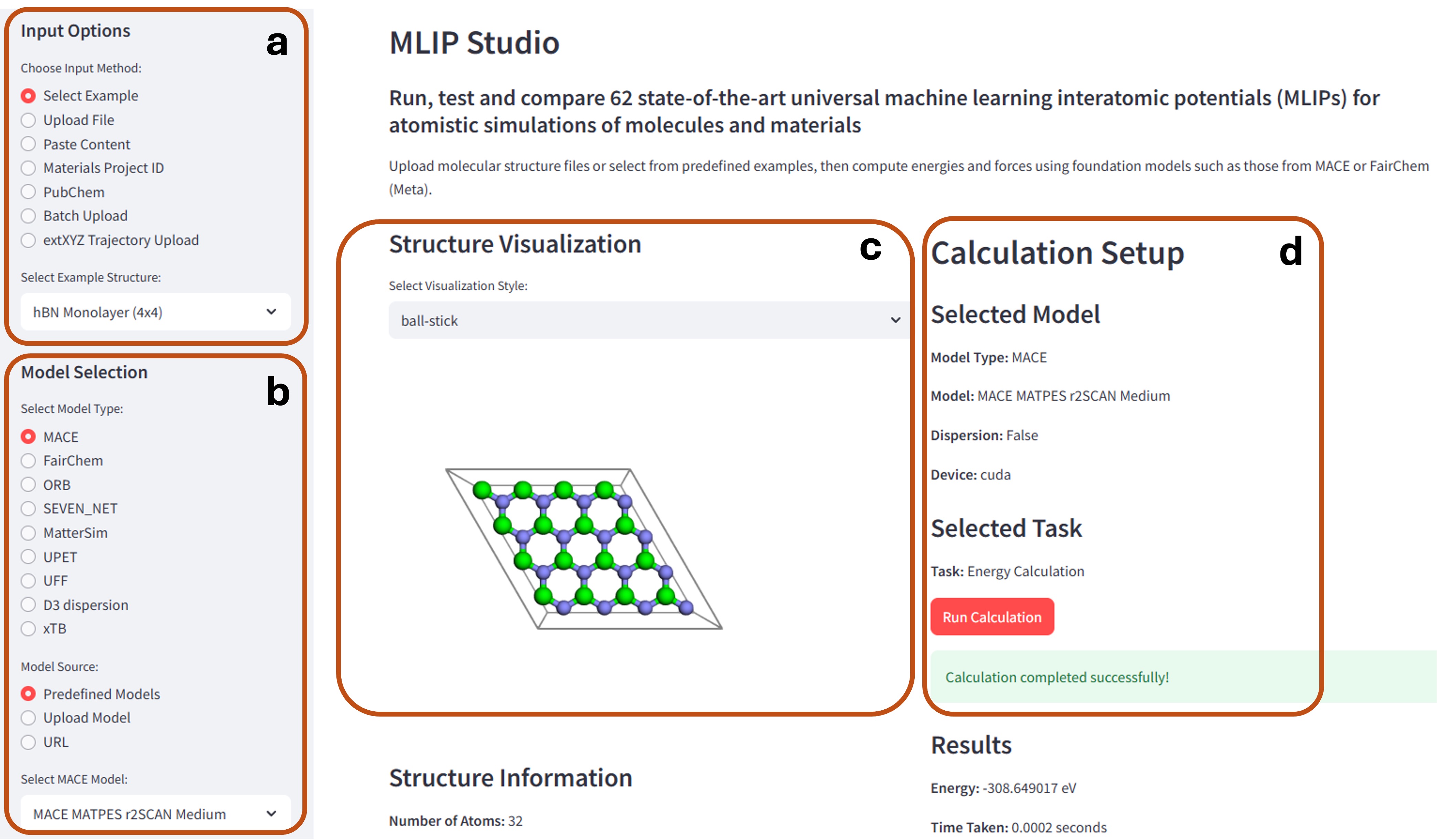}
    \caption{Annotated screenshot of the MLIP Studio home page. 
    (a) Input options panel showing the available methods for providing a chemical structure: selecting a built-in example, uploading a file, pasting file contents, importing from Materials Project or PubChem, batch uploading multiple files, or uploading an extXYZ trajectory. 
    (b) Model selection panel listing the supported MLIP families, semi-empirical and classical calculators, and options for using predefined models, uploading a custom model, or specifying a model URL. 
    (c) Interactive three-dimensional structure visualization of the selected or imported structure (here, a $4\times4$ hBN monolayer), rendered using Py3Dmol. 
    (d) Calculation setup panel displaying the selected model type and name, dispersion correction status, computational device, selected task, and the ``Run Calculation'' button.}
    \label{fig:homepage}
\end{figure}

The calculation setup panel, shown in Figure~\ref{fig:homepage}d, summarizes the current configuration before execution. 
It displays the selected model type (e.g., MACE), the specific model name (e.g., \texttt{MACE MPA Medium}), whether dispersion correction is enabled (applicable to MACE models that support D3 corrections~\cite{grimme2011effect}), and the computational device (CPU or CUDA, if a compatible GPU is available). 
The panel also shows the selected calculation task. Once all parameters are configured, the user initiates the computation by clicking the ``Run Calculation'' button. 
Upon completion, results are displayed directly in the browser, as described in Section~\ref{sec:features}.

\section{Features and Applications}
\label{sec:features}

MLIP Studio supports a diverse set of atomistic calculation tasks, each accessible through the task selection menu in the calculation setup panel (Figure~\ref{fig:homepage}d). 
In this section, we describe each available task, the corresponding output, and illustrative applications.

\subsection{Energy, Force, and Stress Evaluation}
\label{sec:efs}

The most fundamental task is the single-point evaluation of the total energy, atomic forces, and stress tensor for a given structure. 
Upon completion, the application reports the total potential energy (in eV), the maximum atomic force magnitude (in eV/\AA{}), the individual force components on each atom in a downloadable table, and the full $3\times3$ stress tensor. The wall time for the evaluation is also displayed. 
Figure~\ref{fig:efs_atomization}a shows representative output for a water molecule, including the tabulated per-atom force components.

\subsection{Atomization and Cohesive Energy}
\label{sec:atomization}

\begin{figure}[htbp]
    \centering
    \includegraphics[width=\textwidth]{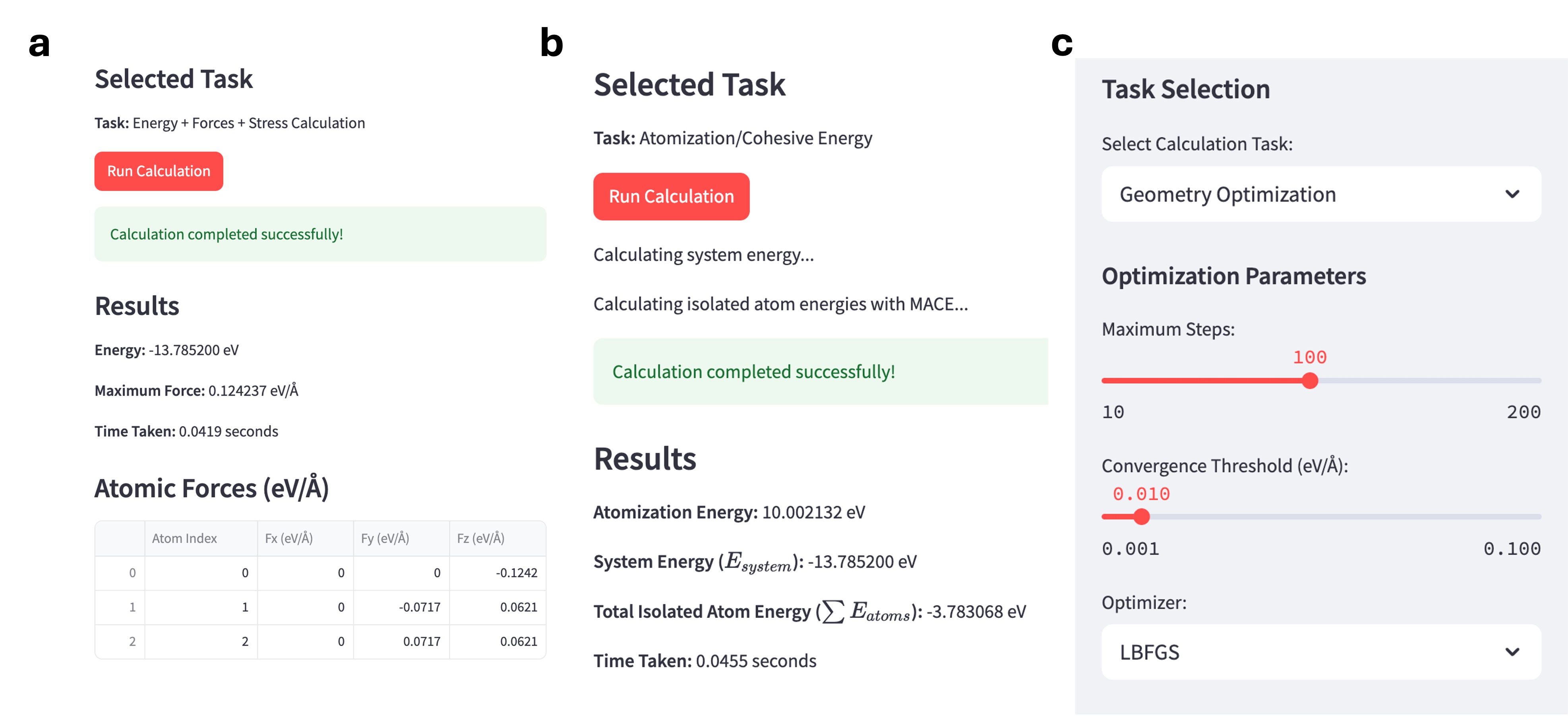}
    \caption{Screenshots of MLIP Studio output for representative calculations. 
    (a) Energy, force, and stress evaluation results for a water molecule, showing the total energy, maximum force, time taken, and the tabulated per-atom force components. 
    (b) Atomization/cohesive energy calculation results using the MACE MPA Medium model, reporting the atomization energy, system energy, and total isolated atom energy. 
    (c) Geometry optimization input settings showing the task selection, maximum number of steps, force convergence threshold, and optimizer selection (here, LBFGS).}
    \label{fig:efs_atomization}
\end{figure}

The atomization/cohesive energy task computes the energy required to dissociate a system into its constituent isolated atoms. 
For molecular (nonperiodic) systems, the quantity reported is the atomization energy; for periodic systems, it is the cohesive energy. 
In both cases, the definition is
\begin{equation}
    E_{\mathrm{at/coh}} = \sum_{i} E_{\mathrm{atom},i} - E_{\mathrm{system}},
    \label{eq:atomization}
\end{equation}
where $E_{\mathrm{system}}$ is the total energy of the system and $E_{\mathrm{atom},i}$ is the energy of each isolated constituent atom computed with the same MLIP. 
The application reports $E_{\mathrm{at/coh}}$, along with $E_{\mathrm{system}}$ and $\sum_i E_{\mathrm{atom},i}$ individually. 
This provides a rapid means of comparing the relative stability of different compounds or polymorphs without requiring a full DFT calculation. 
Figure~\ref{fig:efs_atomization}b shows the result of a cohesive energy calculation performed using the \texttt{MACE MPA Medium} model.

\subsection{Geometry Optimization}
\label{sec:geomopt}

Geometry optimization relaxes the atomic coordinates (and, optionally, the unit cell for periodic systems) to minimize the total potential energy. 
MLIP Studio offers a selection of optimizers available through ASE~\cite{hjorth2017atomic}, including LBFGS~\cite{nocedal1980updating}, BFGS~\cite{broyden1970convergence, fletcher1970new, goldfarb1970family, shanno1970conditioning}, LBFGSLineSearch~\cite{nocedal2006numerical}, BFGSLineSearch~\cite{nocedal2006numerical}, FIRE \cite{bitzek2006structural}, GPMin~\cite{garijo2019local}, and MDMin~\cite{}. 
Users can specify the maximum number of optimization steps and the force convergence criterion (in eV/\AA{}) through the interface, as shown in Figure~\ref{fig:efs_atomization}c.

\begin{figure}[htbp]
    \centering
    \includegraphics[width=\textwidth]{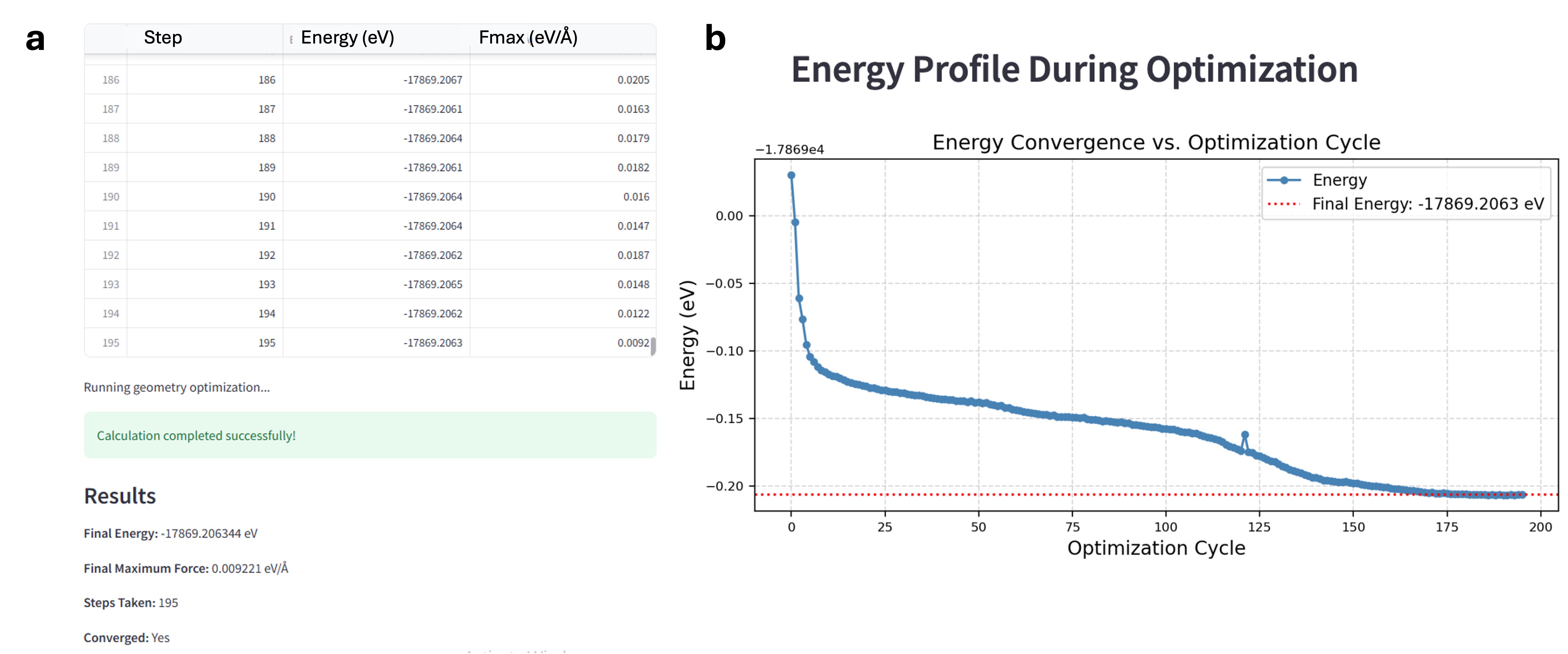}
    \caption{Geometry optimization output for ibuprofen using the UMA OMOL s1.1 model and the LBFGS optimizer with a force convergence threshold of 0.01~eV/\AA{}. (a) Tabulated energy and maximum force ($f_{\mathrm{max}}$) for the final optimization steps, showing convergence in 195 iterations. The final energy, maximum force, number of steps, and convergence status are reported below the table. (b) Energy convergence profile plotted as a function of the optimization cycle, with the final converged energy indicated by a dashed red line.}
    \label{fig:geomopt_output}
\end{figure}

During the optimization, the application displays a live-updated table of the energy and maximum force at each iteration. 
Upon convergence, the following outputs are provided: (i) the final optimized energy and maximum residual force, (ii) the total number of steps taken and whether convergence was achieved, (iii) the optimized geometry, visualized interactively, (iv) a downloadable table of per-step energies and forces, (v) a plot of the energy as a function of the optimization cycle, and (vi) the full optimization trajectory, which can be viewed and downloaded. 
Figure~\ref{fig:geomopt_output}a shows the tabulated per-step optimization data, and Figure~\ref{fig:geomopt_output}b shows the energy convergence plot for an ibuprofen molecule optimized using the \texttt{UMA OMOL s1.1} model with the LBFGS optimizer. 
These comprehensive outputs also make MLIP Studio a valuable pedagogical tool for introducing geometry optimization concepts and comparing the behavior of different optimization algorithms in lecture settings.

In addition to the standard ASE optimizers, we have implemented a custom multi-stage optimizer called Fast-MSO (Fast Multi-Stage Optimizer), which chains three optimizers in a deterministic, monotonic sequence: FIRE $\rightarrow$ MDMin $\rightarrow$ LBFGS. 
The rationale is that FIRE is robust for configurations with large residual forces, MDMin provides efficient downhill relaxation at intermediate force magnitudes, and LBFGS achieves rapid final convergence near the minimum. 
Stage transitions occur when the maximum force (absolute) drops below user-configurable thresholds ($f_{\mathrm{FIRE}}$ and $f_{\mathrm{MDMin}}$, with default values of 0.8 and 0.25~eV/\AA{}, respectively). 
Upon each transition, the new optimizer is freshly initialized to avoid contamination from any stale internal states (e.g., Hessian approximations or velocity vectors) accumulated by the previous optimizer. 
Once the transition happens, it is not reversible even if the maximum force increases at some future step. 
This is done to ensure reproducible behavior.

\subsubsection{Application: MLIP Pre-optimization for Accelerating DFT Calculations}
\label{sec:preopt}

A particularly impactful application of the geometry optimization feature is the pre-optimization of structures prior to DFT calculations. 
In conventional computational workflows, initial geometries constructed using software such as Materials Studio~\cite{meunier2021materials} or Avogadro~\cite{hanwell2012avogadro} are often pre-relaxed with UFF~\cite{rappe1992uff} before being submitted to DFT optimization. 
However, the accuracy of UFF is inferior to that of modern universal MLIPs, resulting in pre-optimized geometries that still require substantial DFT effort to converge. 
By contrast, MLIP pre-optimization yields structures that are already close to the DFT potential energy minimum, dramatically reducing the number of subsequent DFT optimization steps.

\begin{figure}[htbp]
    \centering
    \includegraphics[width=\textwidth]{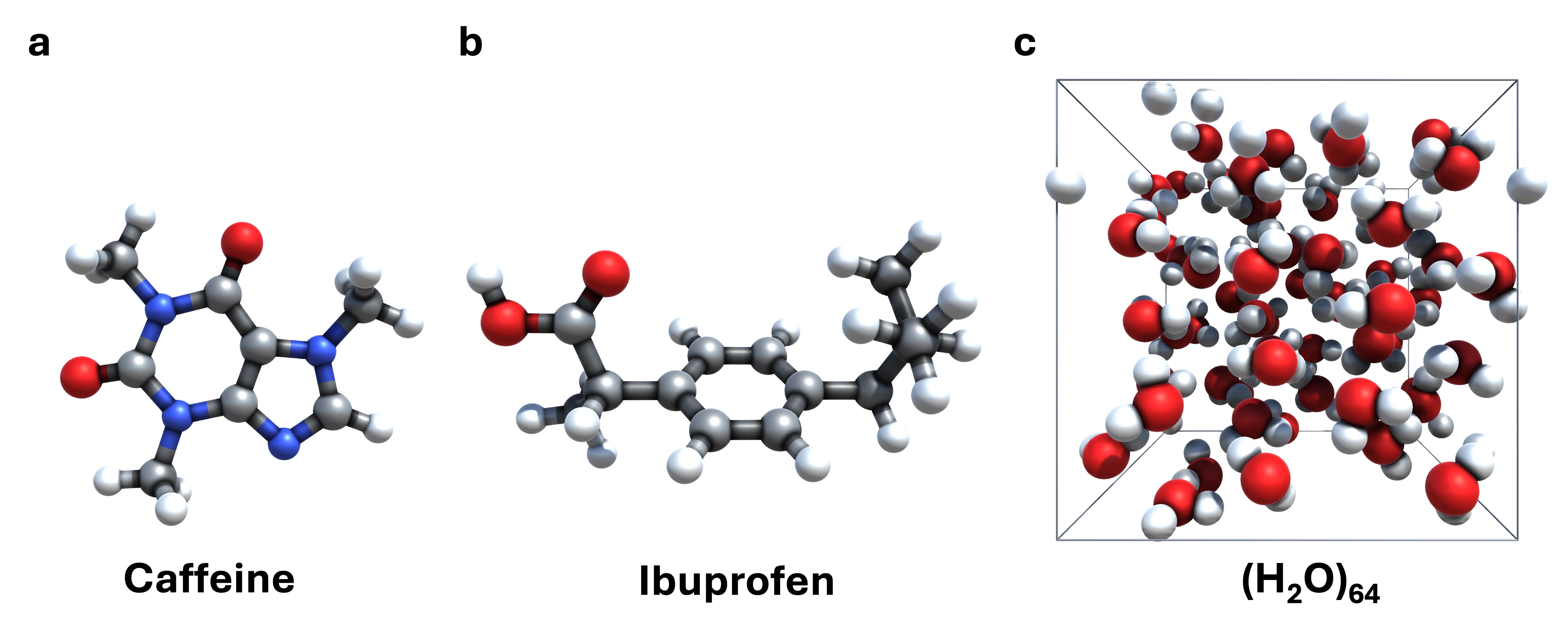}
    \caption{Structures used for the MLIP pre-optimization study. 
    (a) Caffeine molecule. (b) Ibuprofen molecule. 
    (c) Periodic simulation box containing 64 water molecules, $(\mathrm{H_2O})_{64}$, generated using Packmol~\cite{martinez2009packmol}.
    Visualizations generated using CrysX-3D Viewer~\cite{crysx_sharma_2019}.}
    \label{fig:preopt_structures}
\end{figure}

To demonstrate this, we pre-optimized three representative systems using MLIP Studio and subsequently performed full DFT geometry optimizations, comparing the results with DFT optimizations starting from the same initial (non-pre-optimized) geometries. 
The three systems: caffeine, ibuprofen, and a periodic box of 64 water molecules are shown in Figure~\ref{fig:preopt_structures}. 
The MLIP pre-optimization convergence criterion was set to 0.01~eV/\AA{} for the maximum force in all cases.

For caffeine and ibuprofen, pre-optimization was performed using the \texttt{UMA OMOL s1.1} model, which is trained on the OMOL dataset generated with the $\omega$B97M-V functional~\cite{mardirossian2016omegab97m}. 
The subsequent DFT optimizations were carried out using TURBOMOLE v7.9~\cite{franzke2023turbomole,sharma2025density} with the def2-TZVP basis set~\cite{weigend2005balanced}, universal auxiliary basis set~\cite{weigend2005balanced}, and the $\omega$B97M-V functional (convergence criterion: $10^{-3}$~a.u./Bohr for forces and $10^{-3}$~a.u. for energy), ensuring consistency between the MLIP training data and the DFT reference. 
For comparison with a conventional classical pre-relaxation strategy, the same molecular systems were also pre-optimized using UFF~\cite{rappe1992uff} before the DFT optimization. 
For ibuprofen, the MLIP pre-optimization converged in 195 LBFGS iterations. 
The subsequent DFT optimization required only 7 cycles to converge. 
In contrast, the UFF-pre-optimized geometry required 162 DFT optimization cycles, while DFT optimization starting from the same initial geometry without pre-optimization required 185 cycles. 
Thus, UMA pre-optimization reduces the DFT optimization effort by over $26\times$ relative to the unoptimized starting geometry and by over $23\times$ relative to UFF pre-optimization. 
For caffeine, DFT optimization after UMA pre-optimization converged in just 6 cycles, compared to 30 cycles without pre-optimization and 33 cycles after UFF pre-optimization, yielding a speedup of approximately $5\times$ over the from-scratch DFT optimization.

For the water box, pre-optimization was performed using the \texttt{MACE OMAT-0 Medium} model, which is trained on DFT data obtained with the Perdew-Burke-Ernzerhof (PBE) exchange-correlation functional~\cite{PBE_1996,PBE_1997} at the generalized gradient approximation level of theory. 
The subsequent DFT optimization was performed using the Vienna Ab initio Simulation Package (VASP)~\cite{hafner1997vienna, hafner2008ab} (functional: PBE; force convergence criterion: 0.05~eV/\AA{}; plane-wave cutoff: 450 eV; $\Gamma$ point) using the projector augmented wave (PAW) \cite{PAW_1994, PAW_VASP_1999} potentials. 
With MLIP pre-optimization, the VASP geometry optimization converged in only 24 ionic steps, whereas starting from the unoptimized Packmol geometry required approximately 649 ionic steps, showcasing a reduction of over $27\times$. 
The benefit is even more striking when measured in terms of total electronic self-consistent field (SCF) steps: the pre-optimized run required only 180 total SCF iterations compared to 5{,}939 for the from-scratch optimization, corresponding to a speedup of approximately $33\times$. 
This dramatic improvement arises because the randomly packed Packmol geometries contain substantial atomic overlaps and high initial forces. 
Without pre-optimization, the maximum absolute force at the first VASP ionic step was approximately 1.3~eV/\AA{}, whereas the MLIP pre-optimized structure started with a maximum force of just 0.18~eV/\AA{}, placing it immediately much closer to the local minimum, enabling efficient convergence of the DFT optimizer. 
These results are summarized in Table~\ref{tab:preopt}.

\begin{table}[htbp]
    \centering
    \caption{Comparison of the DFT geometry optimization effort with and without MLIP pre-optimization. For caffeine and ibuprofen, DFT calculations were performed with TURBOMOLE v7.9 ($\omega$B97M-V/def2-TZVP; convergence: $10^{-3}$~a.u./Bohr), and pre-optimization used the UMA OMOL s1.1 model (convergence: 0.01~eV/\AA{}). UFF-pre-optimized starting geometries were also tested for the two molecular systems. For $(\mathrm{H_2O})_{64}$, DFT calculations were performed with VASP (PBE; convergence: 0.05~eV/\AA{}) and pre-optimization used the MACE OMAT-0 Medium model (convergence: 0.01~eV/\AA{}).}
    \label{tab:preopt}
    \begin{tabular}{lccccc}
        \toprule
        System & MLIP Steps & \multicolumn{3}{c}{DFT Optimization Cycles/Ionic Steps} & Speedup \\
        \cmidrule(lr){3-5}
         & & With MLIP pre-opt. & With UFF pre-opt. & Without pre-opt. & \\
        \midrule
        Ibuprofen & 195 & 7 & 162 & 185 & $>26\times$ \\
        Caffeine & 26 & 6 & 33 & 30 & $\sim5\times$ \\
        $(\mathrm{H_2O})_{64}$ & 720 & 24 & -- & 649 & $>27\times$ \\
        \bottomrule
    \end{tabular}
\end{table}

\subsection{Vibrational Mode Analysis}
\label{sec:vibrations}

\begin{figure}[htbp]
    \centering
    \includegraphics[width=\textwidth]{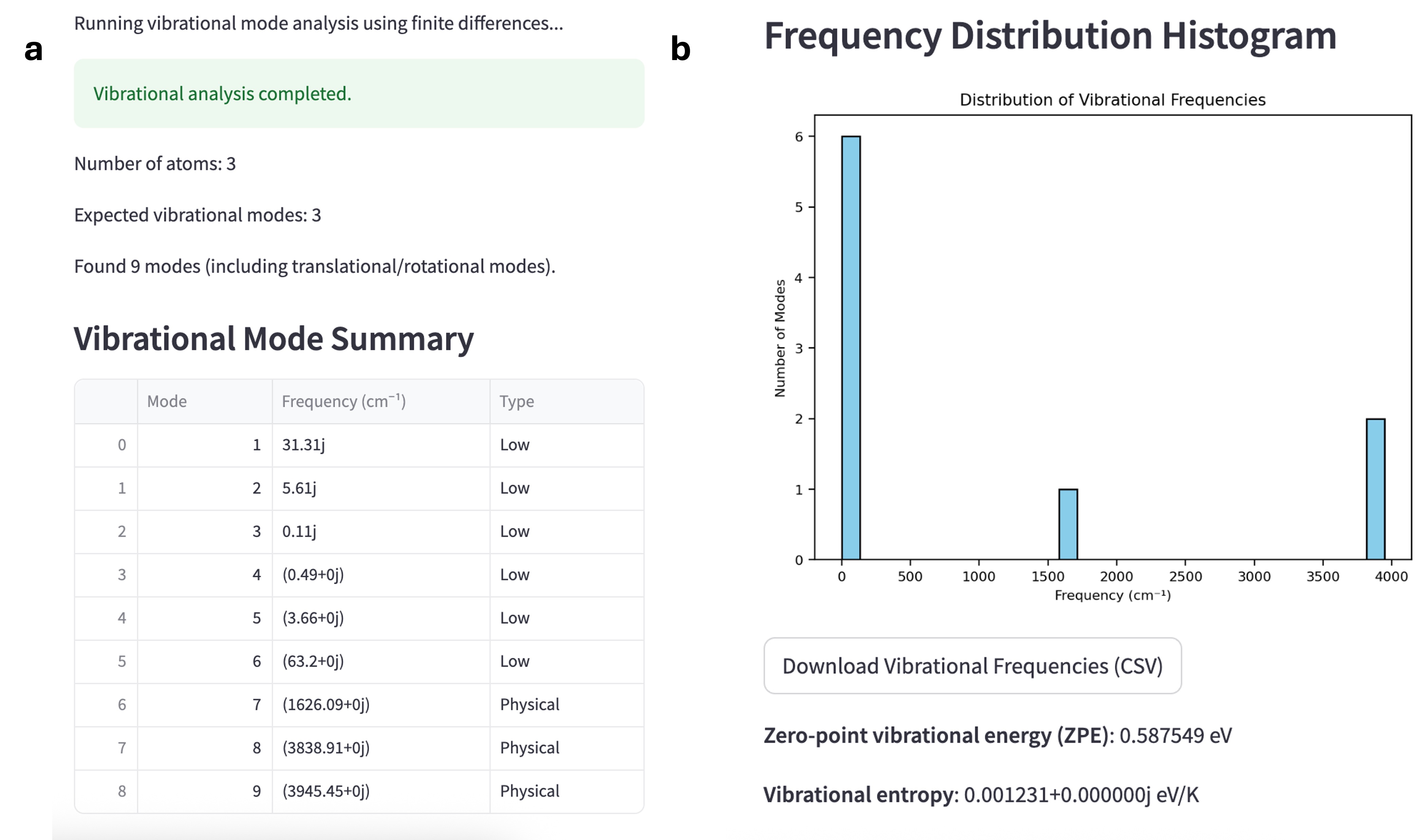}
    \caption{Vibrational mode analysis results for an H$_2$O molecule using the MACE OFF24 Medium model. 
    (a) Vibrational mode summary table listing all nine modes (including translational and rotational modes), their frequencies in cm$^{-1}$, and their classification. 
    Three physical vibrational modes are identified at 1626, 3839, and 3945~cm$^{-1}$. 
    (b) Frequency distribution histogram and the computed zero-point vibrational energy (ZPE) of 0.587~eV.}
    \label{fig:vibrations}
\end{figure}

The vibrational mode analysis task computes the normal-mode frequencies of a structure using finite-difference approximations to the forces. 
Upon completion, the application reports a table of all vibrational modes with their frequencies (in cm$^{-1}$) and classification (physical mode or low-frequency translational/rotational mode), a histogram of the frequency distribution, the zero-point vibrational energy (ZPE) at a user-specified temperature, and a downloadable CSV file of the vibrational data. 
This feature is particularly useful for confirming whether a stationary point obtained, for example, from a nudged elastic band (NEB) calculation, is a true transition state (characterized by exactly one imaginary frequency) or a local minimum. 
Figure~\ref{fig:vibrations} shows the vibrational analysis output for a water molecule computed using the \texttt{MACE OFF24 Medium} model. 
The predicted vibrational modes at 1626, 3839, and 3945~cm$^{-1}$ are in excellent agreement with CCSD(T) values of 1650, 3835, and 3945~cm$^{-1}$, respectively~\cite{howard2014getting}.
Additionally, the predicted ZPE of 0.587~eV is in excellent agreement with the experimental value of 0.584 eV~\cite{csonka2005estimation}.

\subsection{Equation of State}
\label{sec:eos}

The equation of state (EOS) task allows users to compute the energy--volume relationship and extract mechanical properties for periodic systems. 
The user specifies the number of volume points and the volume range as a percentage of the equilibrium volume ($V_0 \pm x\%$), and selects an equation of state for fitting: Birch--Murnaghan~\cite{birch1947finite}, Murnaghan~\cite{murnaghan1944compressibility}, or Vinet~\cite{vinet1986universal}. 
The application uniformly scales the unit cell over the specified volume range, evaluates the energy at each volume using the selected MLIP, and fits the chosen EOS to the resulting $E(V)$ data.

For example, the third-order Birch--Murnaghan equation of state is given by
\begin{equation}
E(V) = E_0 + \frac{9 V_0 B_0}{16} \left\{ \left[ \left(\frac{V_0}{V}\right)^{2/3} - 1 \right]^3 B_0' + \left[ \left(\frac{V_0}{V}\right)^{2/3} - 1 \right]^2 \left[ 6 - 4 \left(\frac{V_0}{V}\right)^{2/3} \right] \right\},
\end{equation}
where $E_0$ is the equilibrium energy (minimum of the $E(V)$ curve), $V_0$ is the equilibrium volume, $B_0$ is the bulk modulus at equilibrium defined as
\begin{equation}
B_0 = V \left. \frac{\partial^2 E}{\partial V^2} \right|_{V=V_0},
\end{equation}
and $B_0'$ is the first pressure derivative of the bulk modulus,
\begin{equation}
B_0' = \left. \frac{\partial B}{\partial P} \right|_{P=0}.
\end{equation}

Here, $B_0$ quantifies the resistance of the material to volume compression, while $B_0'$ describes how this resistance changes under applied pressure.

Similarly, the Murnaghan equation of state is expressed as
\begin{equation}
E(V) = E_0 + \frac{B_0 V}{B_0'(B_0' - 1)} \left[ \left(\frac{V_0}{V}\right)^{B_0'} (B_0' - 1) + 1 \right] - \frac{B_0 V_0}{B_0' - 1},
\end{equation}
and the Vinet equation of state is given by
\begin{equation}
E(V) = E_0 + \frac{2 B_0 V_0}{(B_0' - 1)^2} \left\{ 2 - \left[5 + 3(B_0' - 1)\left(\frac{V}{V_0}\right)^{1/3} - 3\frac{V}{V_0} \right] \exp\left[-\frac{3}{2}(B_0' - 1)\left(\left(\frac{V}{V_0}\right)^{1/3} - 1\right)\right] \right\}.
\end{equation}

\begin{figure}[htbp]
    \centering
    \includegraphics[width=\textwidth]{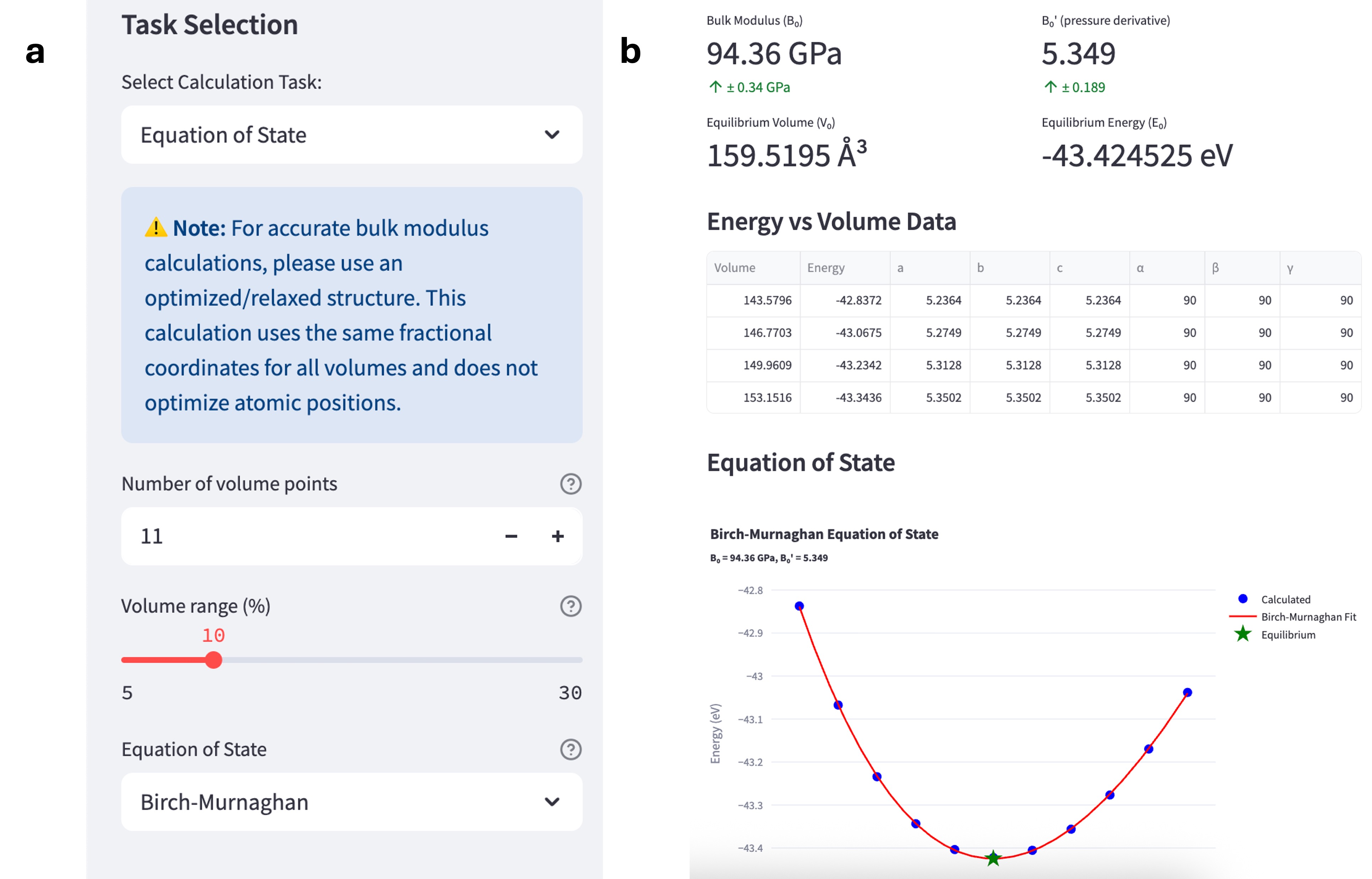}
    \caption{Equation of state calculation for silicon using the MACE OMAT-0 Medium model. (a) Input settings showing the task selection, number of volume points (11), volume range ($\pm 10\%$), and equation of state (Birch--Murnaghan). (b) Output showing the bulk modulus ($B_0 = 94.36$~GPa), pressure derivative ($B_0' = 5.349$), equilibrium volume ($V_0 = 159.52$~\AA$^3$), and equilibrium energy ($E_0 = -43.42$~eV). The energy--volume data table and the fitted Birch--Murnaghan curve (red line) with computed data points (blue circles) and the equilibrium point (green star) are also shown.}
    \label{fig:eos}
\end{figure}

The MLIP Studio outputs include the bulk modulus $B_0$ (in GPa), its pressure derivative $B_0'$, the equilibrium volume $V_0$, and the equilibrium energy $E_0$, all with associated fitting uncertainties. 
A table of energy versus volume data, including the lattice constants at each volume, is provided for downloading. The fitted equation of state is also plotted along with the computed data points, with the equilibrium point marked. 
Figure~\ref{fig:eos} shows both the input settings and the output for a silicon crystal computed using the \texttt{MACE OMAT-0 Medium} model with the Birch--Murnaghan EOS.

\subsection{Spin State Determination}
\label{sec:spin}

The spin determination task leverages the ability of certain MLIPs, notably the UMA OMOL family, trained on the OMOL dataset that includes spin and charge information, to accept spin multiplicity and charge as input parameters. 
By scanning over a range of spin multiplicities (i.e., numbers of unpaired electrons) for a given molecule, the application identifies the ground-state spin as the configuration with the lowest total energy. 
This is directly applicable to providing a good guess value of the magnetic moment (\texttt{MAGMOM} in VASP) or the fixed number of unpaired electrons (\texttt{NUPDOWN} in VASP) to specify in subsequent VASP DFT calculations.

\begin{figure}[h]
    \centering
    \includegraphics[width=\textwidth]{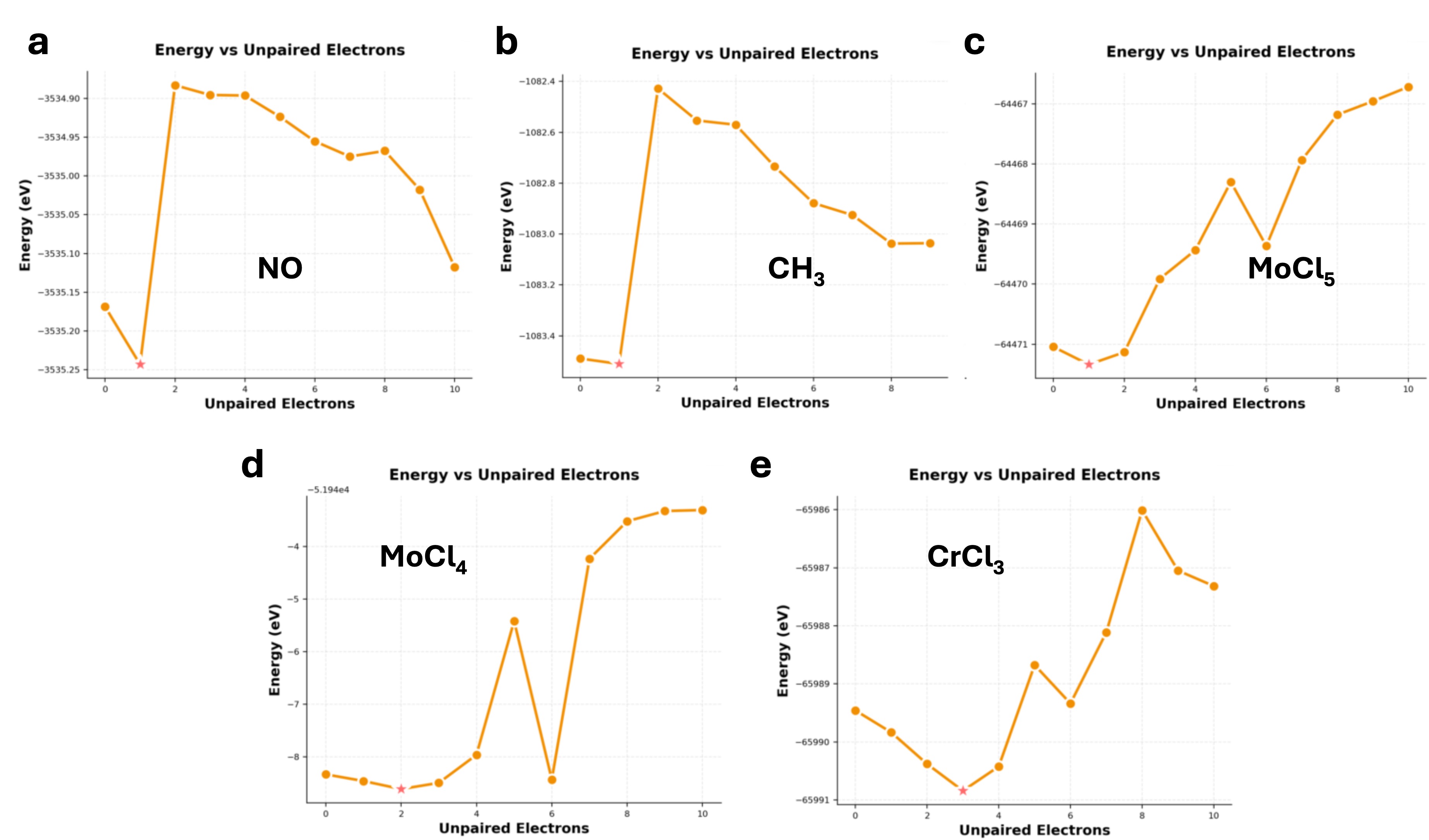}
    \caption{Spin-state determination using the UMA OMOL s1.1 model. Energy as a function of the number of unpaired electrons for (a) NO, (b) CH$_3$, (c) MoCl$_5$, (d) MoCl$_4$, and (e) CrCl$_3$. In each panel, the red star marks the energy minimum, corresponding to the predicted ground-state spin. The UMA OMOL s1.1 model correctly identifies the ground state as a doublet for NO, CH$_3$, and MoCl$_5$; a triplet for MoCl$_4$; and a quartet for CrCl$_3$, in agreement with expected values.}
    \label{fig:spin}
\end{figure}

We validated this feature on five molecules with known ground-state spin multiplicities: NO (doublet, 1 unpaired electron), CH$_3$ (doublet, 1 unpaired electron), MoCl$_5$ (doublet, 1 unpaired electron), MoCl$_4$ (triplet, 2 unpaired electrons), and CrCl$_3$ (quartet, 3 unpaired electrons). 
In all cases, the \texttt{UMA OMOL s1.1} model correctly identified the ground-state spin, as shown by the energy vs.\ number of unpaired electrons plots in Figure~\ref{fig:spin}. 
The energy minimum in each plot corresponds to the known spin ground state, confirming the physical reliability of the spin-dependent energy predictions.

\subsection{Batch Processing and Trajectory Analysis}
\label{sec:batch}

MLIP Studio provides two closely related bulk-processing modes that enable efficient evaluation of large numbers of configurations: the batch uploader and the extXYZ trajectory uploader.

\subsubsection{Batch Uploader}
\label{sec:batch_uploader}

The batch uploader allows users to upload multiple structure files simultaneously in any combination of the supported formats (CIF, extXYZ, XYZ, POSCAR, TURBOMOLE, MOL). 
The application evaluates the energy, forces, and stress tensor for each structure using the selected MLIP. 
Upon completion, it reports a summary table listing the energy, maximum force, and stress of each configuration, histograms showing the distributions of energies, forces, and stresses across all configurations, and a plot of the potential energy as a function of the configuration index.

\subsubsection{extXYZ Trajectory Uploader} 
\label{sec:trajectory}

The extXYZ trajectory uploader accepts a single extXYZ file containing multiple frames and evaluates each frame sequentially. 
This mode has two principal applications:

\paragraph{Potential Energy Surface Screening.}
In many computational workflows, simulation cells must be initialized with randomly packed molecules.
For example, a box of water molecules generated by Packmol~\cite{martinez2009packmol}. 
Since Packmol generates configurations stochastically, the resulting structures can vary significantly in energy, and selecting a low-energy initial configuration is desirable for subsequent molecular dynamics simulations. 
MLIP Studio enables rapid screening of such configuration ensembles. 
As a demonstration, we uploaded 1{,}000 configurations of a periodic box containing 64 water molecules and evaluated the potential energy of each using the \texttt{MACE OMAT-0 Medium} model. 
The energies ranged from $-913$~eV (most stable) to $-904$~eV (least stable), a spread of $\sim$9~eV. 
Using the energy profile and the downloadable tabulated data, the lowest-energy configuration can be readily identified and extracted for further simulations.

\begin{figure}[h]
    \centering
    \includegraphics[width=\textwidth]{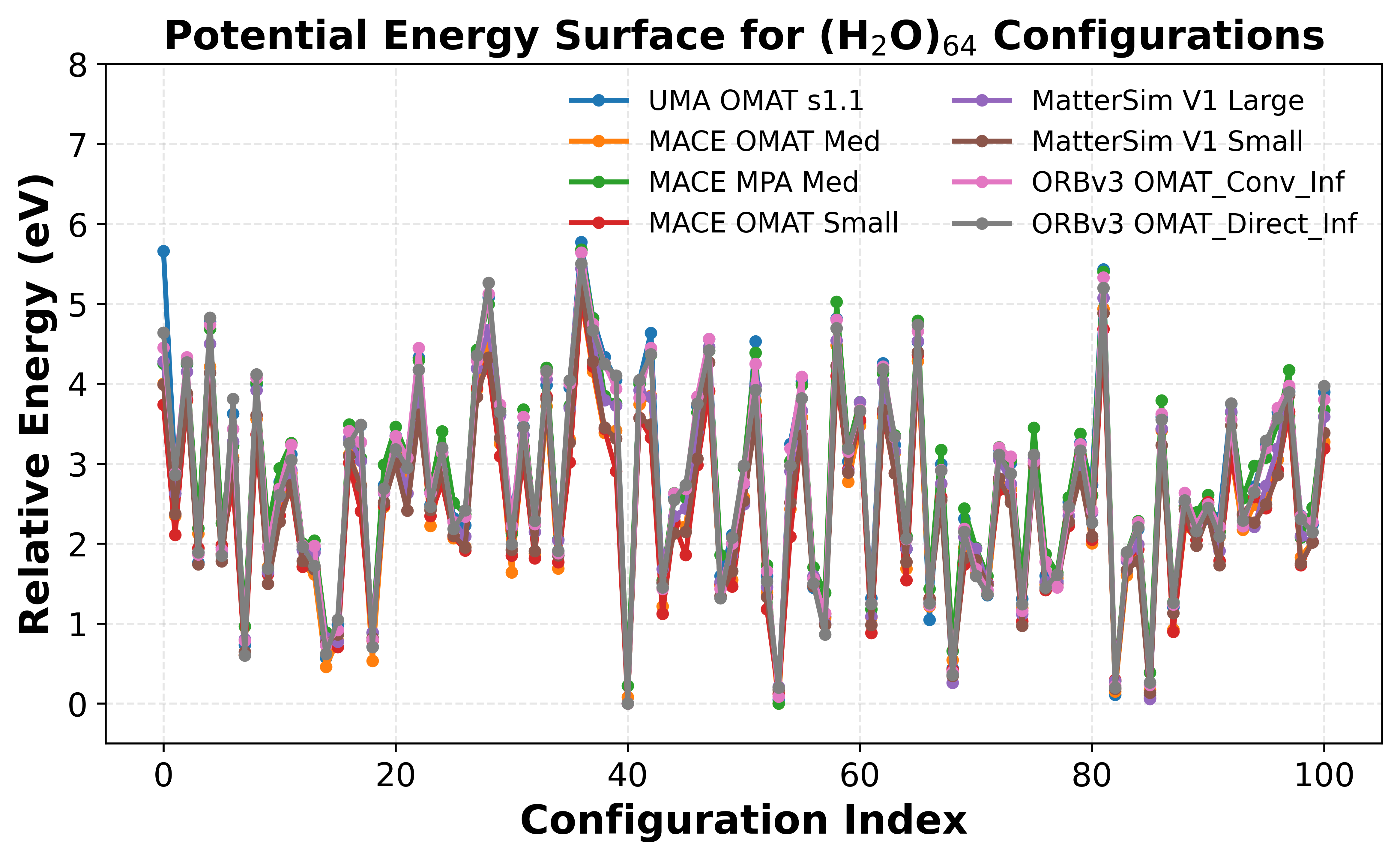}
    \caption{Potential energy profiles for the first 100 of 1{,}000 randomly packed $(\mathrm{H_2O})_{64}$ configurations, evaluated using eight different universal MLIPs trained on PBE-level data: UMA OMAT s1.1, MACE OMAT-0 Medium, MACE MPA Medium, MACE OMAT Small, MatterSim V1 Large, MatterSim V1 Small, ORBv3 OMAT Conservative ($\infty$), and ORBv3 OMAT Direct ($\infty$). All models produce qualitatively consistent energy landscapes and agree on the identification of the highest- and lowest-energy configurations.}
    \label{fig:pes_comparison}
\end{figure}

This trajectory analysis also provides an opportunity to assess the consistency of predictions across different universal MLIPs, a critical consideration for establishing confidence in MLIP-based workflows. 
If models trained on similar data with the same level of theory (e.g., the PBE functional) disagree substantially on the relative energetics, the reliability of their predictions would be called into question. 
We therefore evaluated the 1{,}000 water-box configurations using eight different MLIPs: \texttt{UMA OMAT s1.1}, \texttt{MACE OMAT-0 Medium}, \texttt{MACE MPA-0 Medium}, \texttt{MACE OMAT Small}, \texttt{MatterSim V1 Large}, \texttt{MatterSim V1 Small}, \texttt{ORBv3 OMAT Conservative ($\infty$)}, and \texttt{ORBv3 OMAT Direct ($\infty$)}. 
Figure~\ref{fig:pes_comparison} shows the PES for the first 100 of the 1{,}000 total configurations so as to be able to see a clear distinction between the models. 
All models produce qualitatively consistent potential energy surfaces, agreeing on the relative ordering of configurations and identifying the same configurations as the highest- and lowest-energy structures. 
Similarly, MACE OMOL and UMA OMOL models, both trained on the OMOL dataset, also show excellent mutual agreement. 
This cross-model consistency is encouraging and supports the use of universal MLIPs as reliable tools for configuration screening.

\paragraph{Benchmarking Against Reference Data.}
If the input extXYZ file contains reference energies, forces, and/or stresses (e.g., from DFT calculations), the trajectory uploader automatically generates parity plots comparing the MLIP predictions with the reference data. 
For energies and stresses, scalar parity plots are produced; for forces, an atom-decomposed parity plot colored by element is generated, enabling identification of which atomic species exhibit the largest prediction errors. 
Error metrics including the mean absolute error (MAE), root-mean-square error (RMSE), and coefficient of determination ($R^2$) are reported.

\begin{figure}[h]
    \centering
    \includegraphics[width=\textwidth]{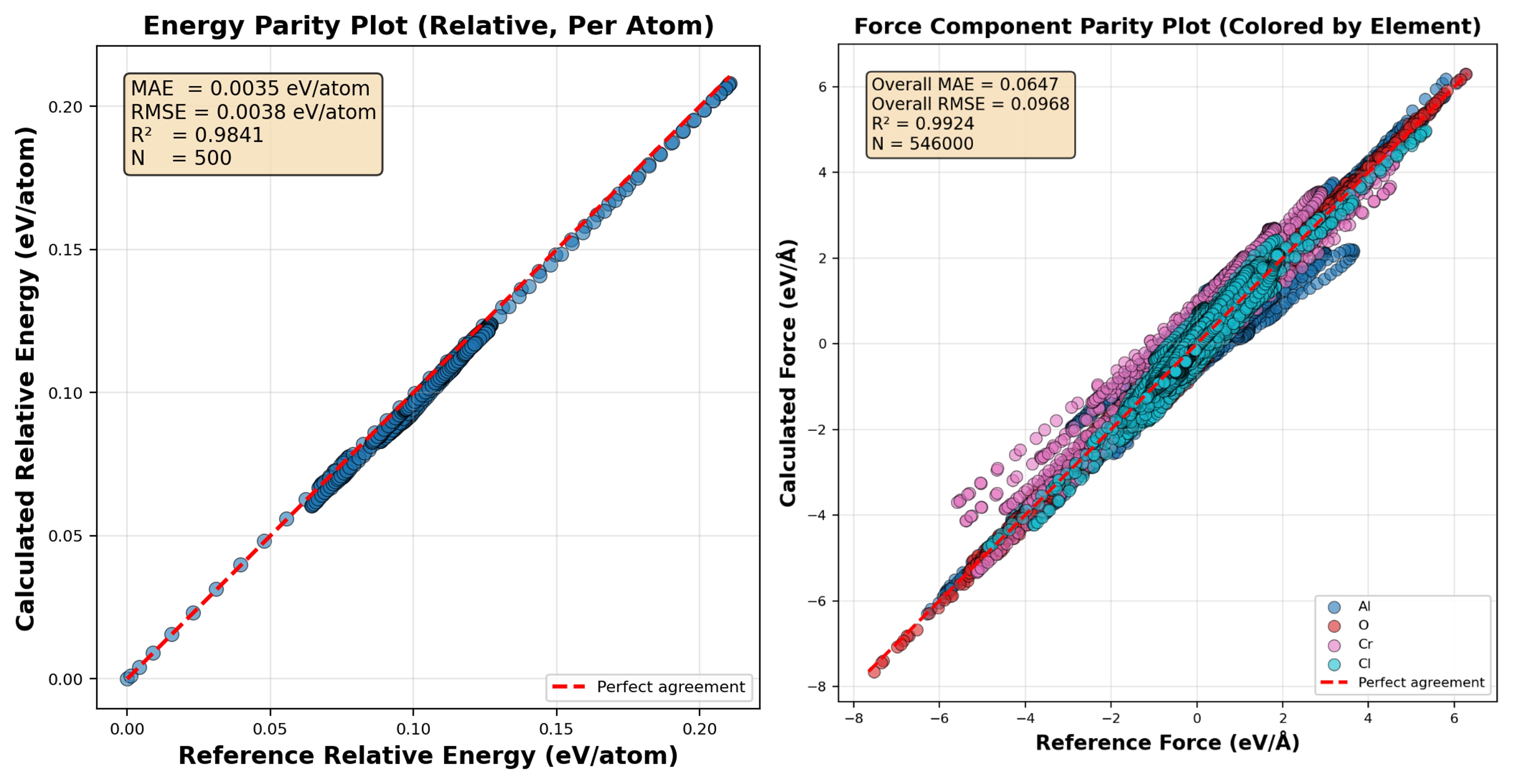}
    \caption{Benchmarking of the MACE OMAT-0 Medium model against DFT reference data for 500 AIMD frames of CrCl$_3$ on a sapphire substrate at 800~K. 
    (Left) Energy parity plot. (Right) Atom-decomposed force component parity plot, colored by element (Al, O, Cr, Cl). 
    The overall MAE is 0.065~eV/\AA{}, RMSE is 0.097~eV/\AA{}, and $R^2 = 0.992$ over $N = 546{,}000$ force components. The dashed red line indicates perfect agreement.}
    \label{fig:parity}
\end{figure}

As a demonstration, we uploaded a trajectory of 500 frames from an ab initio molecular dynamics (AIMD) simulation of a CrCl$_3$ molecule on a sapphire (Al$_2$O$_3$) substrate, performed in ref \citenum{kumar2025tailored} by the authors using VASP at 800~K . 
The \texttt{MACE OMAT-0 Medium} model was used to predict the energies and forces for all 500 frames. 
The resulting force parity plot, shown in Figure~\ref{fig:parity}, demonstrates good agreement between the MLIP predictions and the DFT reference, with an overall force RMSE of 0.097~eV/\AA{} and $R^2 = 0.992$. 
The element-resolved coloring reveals that Cr atoms exhibit slightly larger force errors compared to Al, O, and Cl atoms, providing actionable diagnostic information for potential model refinement.
Additionally, MLIP Studio also displays a downloadable table with reference and predicted values as well as the errors for each frame in the uploaded trajectory.
It is also possible to sort the table by descending values of errors.
This is extremely crucial to identify the frames with larger deviations from the reference values and enables quick diagnostics. 

\subsection{Custom Model Upload}
\label{sec:custom_model}

The rapidly evolving MLIP landscape means that new MACE-based models are being released at an accelerating pace by diverse research groups. 
To provide maximum flexibility, MLIP Studio allows users to supply their own MLIP with MACE architecture, either by uploading a model file directly to the web application or by providing a URL from which the model can be downloaded. 
The uploaded custom model can then be used for all available inference tasks: energy evaluation, geometry optimization, vibrational analysis, equation of state, and so on, within the same unified interface.

This capability is particularly powerful when combined with the trajectory uploader with reference data (Section~\ref{sec:trajectory}). 
Users can upload a custom finetuned model and immediately benchmark it against their DFT reference dataset, obtaining parity plots and error metrics without writing any code. 
We anticipate that this feature will be highly useful in model development workflows, where rapid diagnostic evaluation and visual comparison of model performance are essential for guiding iterative refinement.

\subsection{Computational Performance Benchmarking}
\label{sec:benchmarking}

All calculations in MLIP Studio report the wall-clock time upon task completion, enabling systematic benchmarking of MLIP inference speed. 
This is an important practical consideration, since model selection involves a trade-off between accuracy and computational cost, and the relative performance of different models can vary substantially depending on the hardware.

To provide indicative performance comparisons, we benchmarked the energy, force, and stress inference speed of representative MLIPs by evaluating 1{,}000 configurations of a periodic box containing 64 water molecules (the same configurations discussed in Section~\ref{sec:trajectory}; structure files are provided in the Supporting Information). 
All simulations used \texttt{float32} numerical precision. For ORBv3, we used the \texttt{float32-high} precision setting, which was the fastest available option.

\subsubsection{CPU Benchmarks}

CPU benchmarks were performed on an Intel Core i9-12900K processor (12th generation) with 32~GB memory running Windows 11. 
The results, shown in Figure~\ref{fig:benchmarks} (top panel), reveal a wide range of inference speeds. 
\texttt{PET-MAD XS v1.5 Direct} is the fastest model taking only 72~s, which is impressively fast compared to other models.
\texttt{MatterSim V1 Small} and \texttt{MACE OMAT-0 Small} are the next-fastest models, requiring around 329~s and 409~s, respectively, for 1{,}000 evaluations. 
Despite its larger parameter count (5M parameters), \texttt{MatterSim V1 Large} also performs efficiently at 731~s. 
The medium-sized MACE models: OMAT, MPA, OFF24, and MatPES r2SCAN, all exhibit similar performance, completing 1{,}000 evaluations in approximately 1{,}000~s (i.e., $\sim$1 configuration per second). 
\texttt{ORBv3 OMAT Direct ($\infty$)} performs slightly faster at 916~s, while \texttt{ORBv3 OMAT Conservative ($\infty$)} is notably slower at 1{,}905~s, taking roughly twice as long as the direct variant. 
The slowest models on CPU are the \texttt{UMA s1.1} models (OMAT and OMOL task heads), each requiring roughly 5{,}400~s, and the \texttt{MACE OMOL XL} model, which required approximately 6{,}800~s for 1{,}000 evaluations.

\subsubsection{GPU Benchmarks}

\begin{figure}[htbp]
    \centering
    \includegraphics[width=\textwidth]{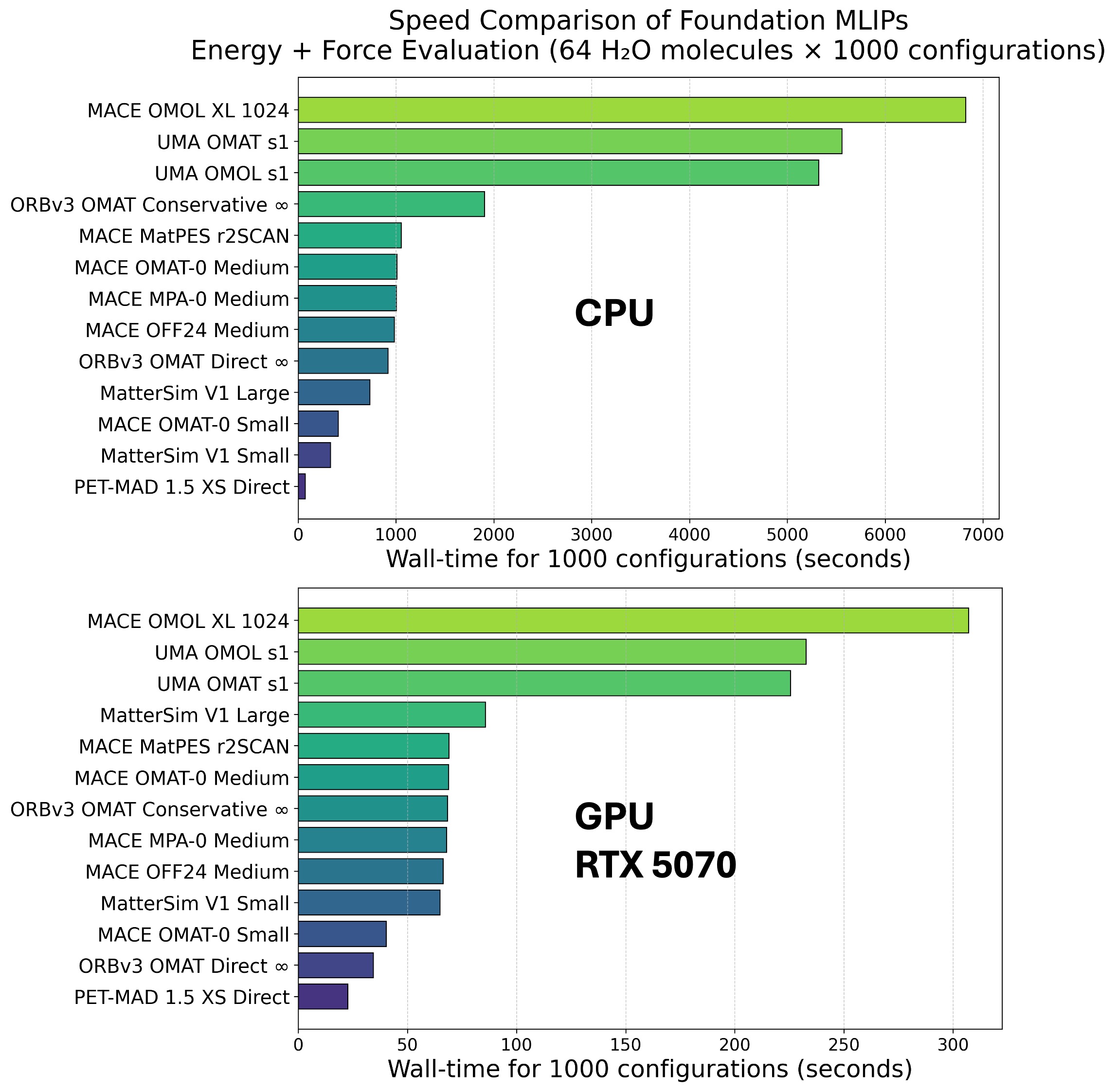}
    \caption{Computational performance benchmarks for universal MLIPs, measured as wall-clock time for 1{,}000 energy, force, and stress evaluations on $(\mathrm{H_2O})_{64}$ configurations. 
    (Top) CPU benchmarks on an Intel Core i9-12900K (32~GB RAM, Windows 11). 
    (Bottom) GPU benchmarks on an NVIDIA GeForce RTX 5070 (12~GB VRAM). 
    All simulations used \texttt{float32} precision. 
    Note the different horizontal axis scales between the two panels.}
    \label{fig:benchmarks}
\end{figure}

GPU benchmarks were performed on an NVIDIA GeForce RTX 5070 (12~GB VRAM; rated FP32 throughput of 27.6~TFLOPS, compared to approximately 0.8~TFLOPS for the i9-12900K CPU). 
The GPU temperature remained below 60$^\circ$C throughout all simulations. 
The GPU results are shown in Figure~\ref{fig:benchmarks} (bottom panel). 

On the GPU, the performance landscape changed considerably compared to the CPU. 
While the \texttt{PET-MAD XS v1.5 Direct} model still remained the fastest on the GPU taking merely $\sim$23~s to evaluate 1000 configurations, the \texttt{ORBv3 OMAT Direct ($\infty$)} was not far behind, emerging as the second-fastest model, completing 1{,}000 evaluations in just 34.4~s (over $26\times$ faster than on CPU). 
\texttt{MACE OMAT-0 Small} was the third fastest at 40.3~s. 
The medium-sized MACE models (OMAT, MPA, OFF24, MatPES r2SCAN) and \texttt{ORBv3 OMAT Conservative ($\infty$)} clustered around 66--69~s. 
Notably, \texttt{MatterSim V1 Small}, which was the second-fastest model on the CPU, lost its advantage on the GPU (65~s), performing comparably to the medium-sized MACE models. 
\texttt{MatterSim V1 Large} similarly did not benefit as much from GPU acceleration (86.0~s) relative to its CPU performance. 
The \texttt{MACE OMOL XL 1024} and UMA models remained the slowest, with the MACE model requiring 307 s and the UMA models taking 225--233~s on GPU, though this still represents a $\sim$$23\times$ speedup over their CPU performance.

Based on these results, we recommend \texttt{PET-MAD XS v1.5}, \texttt{MACE OMAT-0 Small} and \texttt{MatterSim V1 Small} for CPU devices, and \texttt{PET-MAD XS v1.5}, \texttt{ORBv3 OMAT Direct ($\infty$)}, and \texttt{MACE OMAT-0 Small} for GPU devices, when inference speed is the primary consideration. 
However, for more accurate calculations, the MACE medium models are expected to be superior.
These models can be further finetuned on domain-specific data to improve accuracy without sacrificing their speed advantage.

We emphasize that benchmarking the performance of computational codes is inherently sensitive to specific configurations of software libraries, compiler optimizations, and hardware. The results presented here are intended to provide an indicative comparison and should be interpreted accordingly. 
Reproducing the exact timings reported by the original developers of each MLIP would require matching their specific software and hardware environments. 
Nevertheless, the relative performance ratios observed in our benchmarks are physically reasonable and internally consistent, and we believe they provide a qualitatively reliable guide for model selection.

\subsection{Band (HOMO-LUMO) Gap  and Density of States}
\begin{figure}[htbp]
    \centering
    \includegraphics[width=\textwidth]{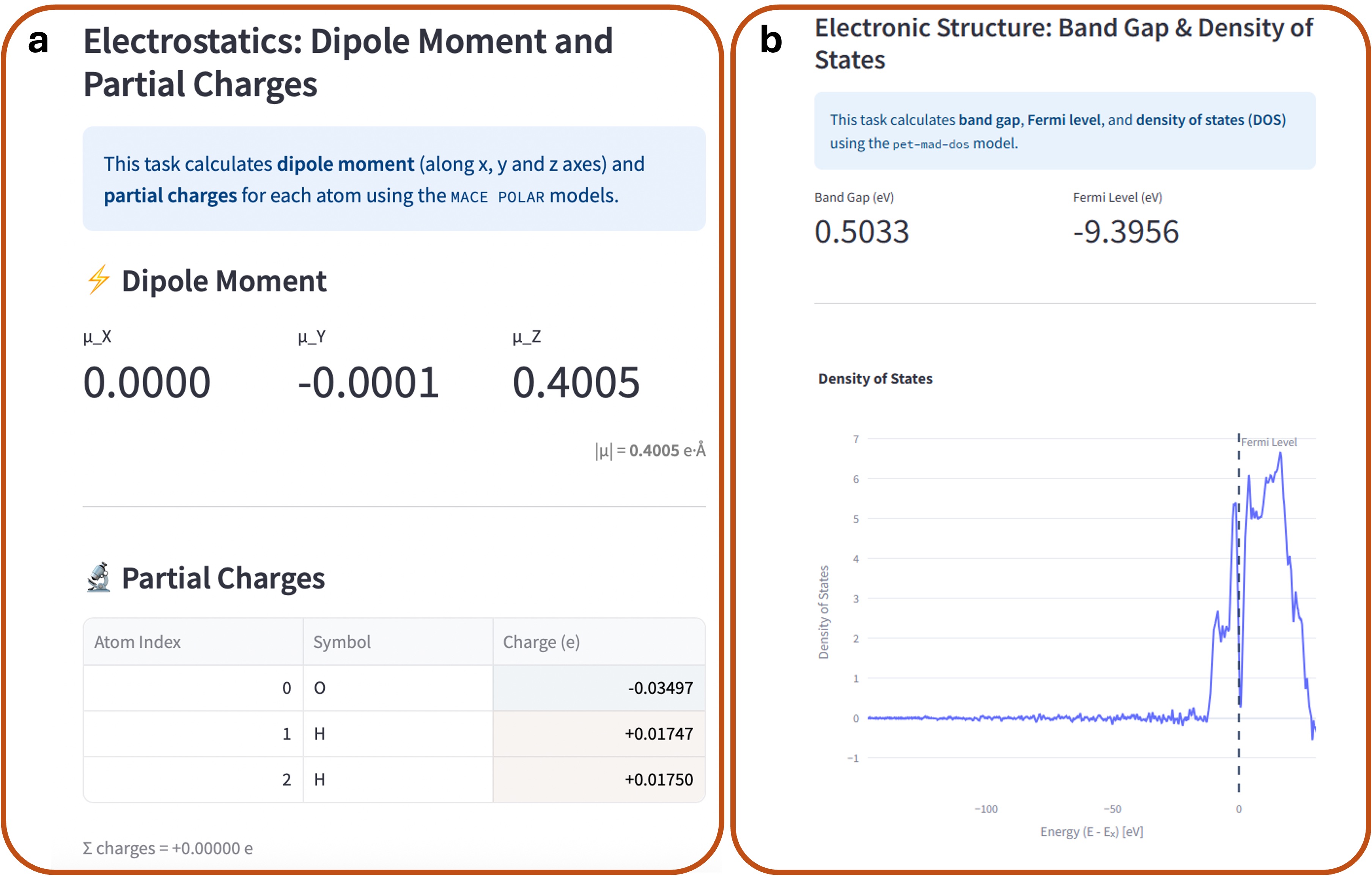}
    \caption{Prediction of dipole moment, atom-resolved partial charges, band gap, and DOS. 
    (a) Output of the dipole moment and partial charges prediction for a water molecule using the \texttt{MACE-POLAR-1 M} model. 
    The model accounts for long-range electrostatics and polarization effects, enabling a consistent description of charge distribution across the system.
    (b) Output of the electronic DOS, Fermi level, and band gap for crystalline silicon using the \texttt{PET-MAD-DOS} model.}
    \label{fig:dipole_band_gap}
\end{figure}

The electronic band gap is the energy difference between the valence band maximum and conduction band minimum, and is a key quantity governing the electrical and optical properties of materials.  For molecules, the electronic energy gap arises from the difference between the energies of the lowest unoccupied molecular orbital (LUMO) and the highest occupied molecular orbital (HOMO).
The DOS describes the distribution of available electronic states as a function of energy and provides a detailed picture of the electronic structure.
The DOS, Fermi level, and band gap can be predicted on MLIP Studio using the \texttt{PET-MAD-DOS} model~\cite{how2026universal}, a universal model for electronic properties of molecules and materials. 
Fig.~\ref{fig:dipole_band_gap}(b) shows how MLIP Studio can be used to predict these electronic properties for the silicon crystal.
Efficient DOS prediction enables rapid screening of materials for electronic and optoelectronic applications, and can be combined with atomistic simulations to estimate finite-temperature electronic properties such as electronic heat capacity.

Additionally, we provide a lightweight, in-house developed model (\texttt{MLIP Studio QM9 Gap}) for predicting HOMO–LUMO gaps of organic molecules. 
The model is a four-layer message-passing graph neural network trained on the QM9 dataset~\cite{ramakrishnan2014quantum}. 
It achieves a test MAE of 0.062~eV, which is comparable to the performance of established architectures such as SchNet~\cite{schutt2018schnet}, PPGN~\cite{maron2019provably}, Cormorant~\cite{anderson2019cormorant}, and MGCN~\cite{lu2019molecular}.

\subsection{Dipole Moment and Partial Charges}
The dipole moment is a measure of the separation of positive and negative charges in a system and provides insight into molecular polarity and electrostatic interactions. Atom-resolved partial charges further quantify the distribution of electronic charge across atoms, enabling a detailed description of charge transfer and local electrostatic environments.
It is possible to compute the dipole moments (along $x$, $y$, $z$) and atom-resolved partial charges on MLIP Studio using the recently released \texttt{MACE-POLAR-1} models~\cite{batatia2026mace}, which extend the MACE framework to include long-range electrostatics and polarization via a non-local, charge-density-based formulation. 
An example for a water molecule is shown in Fig.~\ref{fig:dipole_band_gap}a.
This allows for a physically consistent description of charge redistribution and induction effects across diverse molecular systems.
The predicted partial charges can be used to analyze electrostatic interactions, hydrogen bonding, and charge transfer processes. 
They may also serve as an improved initial guess for electronic structure calculations, for example, by providing environment-dependent atomic charges in superposition of atomic densities (SAD) schemes.

\section{Case Study: End-to-End Workflow for CrCl$_3$ Adsorption on Sapphire}
To illustrate the practical use of MLIP Studio as an end-to-end simulation and benchmarking platform, we consider the CrCl$_3$ molecule and dimer interacting with sapphire ($\alpha$-Al$_2$O$_3$) as a representative molecule--surface interaction case study. This system is important from the perspective of modeling the chemical vapor deposition synthesis of CrCl$_3$, which is a 2D magnetic material.
The authors in ref. \citenum{kumar2025tailored} showed that \texttt{MACE OMAT Medium} performs reasonably well for this system. 
Here, we compare the DFT results from that work with other universal MLIPs (trained on the PBE level DFT data) as facilitated by MLIP Studio.
This system combines several levels of atomistic complexity: a bulk oxide substrate, an isolated transition-metal halide molecule with unpaired electrons, an Al$_2$O$_3$ surface, an adsorbed interface, and a finite-temperature AIMD trajectory. 
It therefore provides a chemically meaningful test case for demonstrating multiple MLIP Studio capabilities within a single workflow.

The case study proceeds in four stages. 
We first validate the bulk Al$_2$O$_3$ substrate by comparing MLIP-predicted lattice parameters, cohesive energies, bulk moduli, and band gaps with DFT or literature reference values. 
We then check the capabilities of universal MLIPs, trained on spin data, to predict the correct spin state of CrCl$_3$. Note that for spin-state determination, we use OMOL-based MLIPs, for which the training data is not at the PBE level of theory but rather at the $\omega$B97M-V level. 
Next, we evaluate the Al$_2$O$_3$ surface energy and the potential energy surface of the CrCl$_3$ dimer on sapphire using several MLIPs and compare the results with DFT reference calculations from ref. \citenum{kumar2025tailored}. 
Finally, we use an AIMD trajectory of the CrCl$_3$/sapphire interface from ref. \citenum{kumar2025tailored} to benchmark the different MLIPs through energy and force error metrics.

The bulk validation results are summarized in Table~\ref{tab:case_bulk}. 
Most MLIPs reproduce the DFT-relaxed hexagonal cell within $\sim$0.4\% for $a$ and $\sim$0.6\% for $c$. 
Interestingly, \texttt{PET-OMAT-V1 M} could not optimize the structure to have forces below the 0.01 eV/\AA threshold utilized here.
The cohesive energies predicted by the MACE and UMA models are clustered near the reported computational and experimental estimates of 6.43~eV/atom~\cite{alavi2003adsorption} and 6.38~eV/atom~\cite{nebergall1968general}, respectively. 
Cohesive energy is an interesting property, as it also checks an MLIP's ability to predict the isolated atomic energies correctly. 
Next, we evaluate the performance for the prediction of bulk modulus.
The EOS-derived bulk moduli reported in literature $\approx$236~GPa~\cite{zhang2016anomalous, golesorkhtabar2013elastic}, are softer than the experimental 252.3~GPa value reported for $\alpha$-Al$_2$O$_3$~\cite{smithells1967metals}. 
Most of the benchmarked MLIPs predict values ($\approx230$ GPa) in good agreement with the reference DFT values, except for the \texttt{MACE MatPES PBE} model. 
For band gap, the \texttt{PET-MAD-DOS} model yields a value of 5.92~eV, which is close to recent PBE-level values near 6~eV, while remaining below the experimental optical gap of $\sim$8.8~eV, as expected for semilocal electronic-structure references~\cite{acharya2025optimizing}.

\begin{table*}[htbp]
\centering
\caption{Bulk $\alpha$-Al$_2$O$_3$ properties from the end-to-end workflow. Lattice parameters are for the hexagonal cell.}
\label{tab:case_bulk}
\begin{adjustbox}{max width=\textwidth}
\begin{tabular}{lccccc}
\toprule
Model/reference & $a$ (\AA) & $c$ (\AA) & $E_{\mathrm{coh}}$ (eV/atom) & $B_0$ (GPa) & $E_g$ (eV) \\
\midrule
DFT, this work & 4.79 & 13.06 & -- & -- & -- \\
Literature/experiment & 4.76~\cite{lee1985structural} & 12.99~\cite{lee1985structural} & 6.38~\cite{nebergall1968general}, 6.43~\cite{alavi2003adsorption} & 236.2~\cite{golesorkhtabar2013elastic}, 252.3~\cite{smithells1967metals} & 6.02 (PBE), 8.8 (exp.)~\cite{acharya2025optimizing} \\
UMA OMAT s1p1 & 4.81 & 13.14 & 6.45 & 235.03 & -- \\
MACE MatPES PBE & 4.79 & 13.07 & 6.44 & 191.80 & -- \\
MACE MPA Medium & 4.80 & 13.12 & 6.46 & 233.95 & -- \\
MACE OMAT Medium & 4.79 & 13.10 & 6.47 & 237.29 & -- \\
ORBv3 OMAT Conv 20 & 4.81 & 13.12 & 5.42 & 222.44 & -- \\
MatterSim V1 S & 4.80 & 13.11 & 6.01 & 233.46 & -- \\
PET-OMAT-V1 M & -- & -- & -- & 231.97 & -- \\
PET-MAD-DOS & -- & -- & -- & -- & 5.92 \\
\bottomrule
\end{tabular}
\end{adjustbox}
\end{table*}

For isolated CrCl$_3$, the molecular spin scan performed using \texttt{MACE POLAR 1 M}, \texttt{UMA OMOL s1p1}, and \texttt{MACE OMOL XL 1024} placed the minimum at three unpaired electrons ($S=3/2$, quartet) for the molecular models used for spin-state determination, consistent with the expected high-spin Cr(III) $d^3$ configuration predicted by DFT. 
This provides a simple but useful check that MLIP Studio can supply a physically sensible spin initialization before more expensive electronic-structure calculations.

\begin{figure}[htbp]
    \centering
    \includegraphics[width=0.78\textwidth]{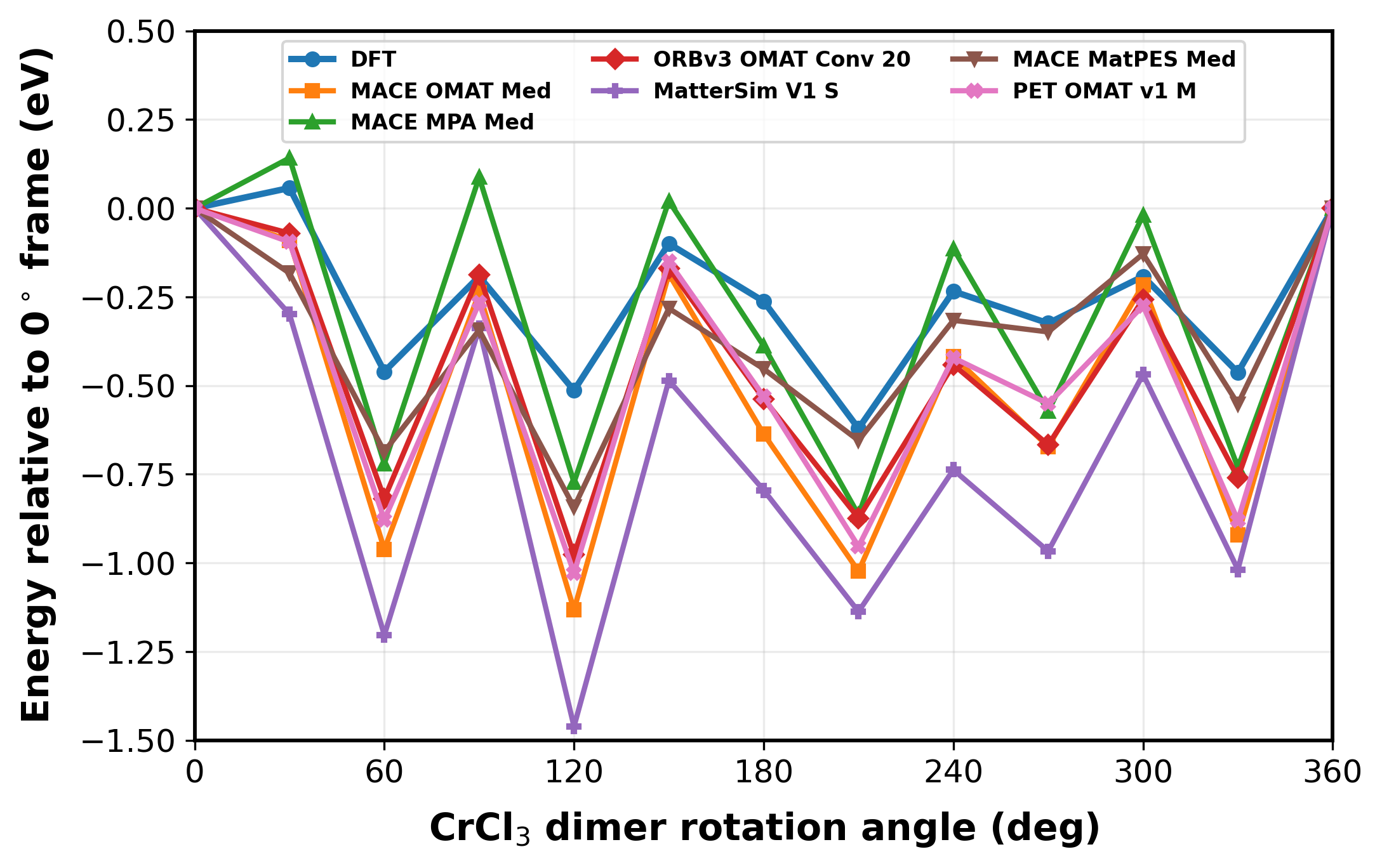}
    \caption{Rotational potential energy curves for a tilted CrCl$_3$ dimer on sapphire. 
    Energy is relative to the 0$^\circ$ configuration.}
    \label{fig:case_pes}
\end{figure}

The surface energies are collected in Table~\ref{tab:case_surface_pes}. 
For the Al-terminated Al$_2$O$_3$(0001) slab, the MLIP surface energies fall between 1.38 and 1.59~J~m$^{-2}$, below but reasonably close to the present DFT value of 1.84~J~m$^{-2}$ and the reported 1.65--2.13~J~m$^{-2}$ range for single-Al-terminated slabs~\cite{ahmad2022modelling}. 
Additionally, we benchmark the MLIPs for the rotational potential energy curve of the tilted CrCl$_3$ dimer on the sapphire surface using the data points from ref. \citenum{kumar2025tailored}.
The potential energy curves obtained from MLIPs are compared with the reference DFT results in Fig.~\ref{fig:case_pes}. 
Most models reproduce the DFT curve qualitatively, with the \texttt{MACE MatPES PBE} curve coming the closest to the reference.
Interestingly, the \texttt{UMA OMAT s1p1} gives a much larger variation of 5.05~eV span and is therefore omitted from the figure for readability.

\begin{table}[htbp]
\centering
\caption{Al-terminated Al$_2$O$_3$(0001) surface energies.}
\label{tab:case_surface_pes}
\begin{adjustbox}{max width=\textwidth}
\begin{tabular}{lc}
\toprule
Model/reference & Surface energy (J~m$^{-2}$)  \\
\midrule
DFT, this work & 1.840  \\
Literature range & 1.65--2.13~\cite{ahmad2022modelling}  \\
UMA OMAT s1p1 & 1.468  \\
MACE MatPES PBE & 1.377  \\
MACE MPA Medium & 1.463  \\
MACE OMAT Medium & 1.595  \\
ORBv3 OMAT Conv 20 & 1.442  \\
MatterSim V1 S & 1.481  \\
\bottomrule
\end{tabular}
\end{adjustbox}
\end{table}

Finally, Table~\ref{tab:case_aimd_metrics} compares MLIP predictions against the 500-frame AIMD trajectory of the CrCl$_3$/sapphire interface obtained from ref. \citenum{kumar2025tailored}. 
Interestingly, \texttt{PET OMAT v1 M} gives the smallest relative-energy error (MAE = 2.2~meV/atom), while \texttt{UMA OMAT s1p1} gives the best force agreement (MAE = 51.8~eV/\AA). 
MatterSim, MACE OMAT, and MACE MPA remain close in force accuracy, whereas ORBv3 shows a much larger relative-energy and force error for this trajectory. 
These results illustrate the main practical value of the workflow: the preferred model depends on the target observable, and MLIP Studio makes that trade-off visible and serves as a useful benchmarking tool.

\begin{table}[htbp]
\centering
\caption{AIMD parity metrics for 500 CrCl$_3$/sapphire frames. Relative energies are per atom and referenced to the first frame.}
\label{tab:case_aimd_metrics}
\begin{adjustbox}{max width=\textwidth}
\begin{tabular}{lcccc}
\toprule
Model & Energy MAE (meV/atom) & Energy RMSE (meV/atom) & Force MAE (meV/\AA) & Force RMSE (meV/\AA) \\
\midrule
PET OMAT M v1 & 2.2 & 2.4 & 71.2 & 103.2 \\
MACE OMAT Medium & 3.5 & 3.8 & 64.3 & 96.7 \\
MACE MPA Medium & 3.3 & 3.4 & 84.7 & 126.2 \\
MatterSim V1 S & 4.3 & 4.6 & 77.9 & 113.7 \\
MACE MatPES PBE & 10.6 & 10.8 & 111.2 & 149.0 \\
UMA OMAT s1p1 & 13.5 & 13.6 & 51.8 & 95.5 \\
ORBv3 OMAT Conv 20 & 429.5 & 536.9 & 165.6 & 283.2 \\
\bottomrule
\end{tabular}
\end{adjustbox}
\end{table}

Together, these illustrative calculations demonstrate how MLIP Studio can be used not only for individual property predictions, but also for constructing a complete workflow for structure optimization, property evaluation, DFT comparison, and model selection on chemically realistic  systems.

\section{Conclusion and Outlook}
\label{sec:conclusion}

In this work, we presented MLIP Studio, a free web-based platform that provides unified access to over 60 universal machine learning interatomic potentials spanning six major model families, including MACE, FairChem (UMA), ORB, MatterSim, and SevenNet, along with semi-empirical (xTB) and classical (UFF) calculators. 
The platform supports a comprehensive suite of atomistic simulation tasks, including single-point energy, force, and stress evaluations; atomization and cohesive energy calculations; geometry optimization with multiple optimizers; vibrational mode analysis; equation-of-state fitting; and spin-state determination. 
Through its batch processing and trajectory analysis capabilities, MLIP Studio enables efficient high-throughput screening of molecular configurations and systematic benchmarking of MLIP predictions against DFT reference data, with automated generation of parity plots and error metrics. 
The ability to upload custom MACE models further extends the platform's utility for model development and diagnostics.
Additionally, built-in wall-time reporting enables systematic benchmarking of model performance across hardware and model families.

We demonstrated several practical applications that highlight the platform's value. 
MLIP-based geometry pre-optimization was shown to accelerate subsequent DFT calculations by factors of $2\times$ to $33\times$ across molecular and periodic systems, offering a compelling alternative to traditional UFF-based pre-relaxation. 
The spin-determination feature was validated against experimentally known ground states for various transition-metal-containing molecules, and is a useful tool for obtaining initial guess values for DFT codes. 
Cross-model comparison of potential energy surfaces for 1{,}000 water-box configurations revealed reassuring qualitative consistency (including identical identification of the lowest- and highest-energy configurations) across eight independently developed MLIPs.
Comprehensive CPU and GPU performance benchmarks provided practical guidance for model selection, revealing that inference speed can vary by over an order of magnitude across models and that relative performance rankings shift between CPU and GPU execution.
We also showed how the bulk evaluation of ensemble configurations can be used to extract the lowest energy structure, as well as the identification of high-error frames when benchmarking against DFT data, enabling better data curation for finetuning and training.
Finally, the CrCl$_3$/sapphire case study demonstrated the use of MLIP Studio as a complete workflow platform for a chemically realistic system, spanning bulk substrate validation, molecular spin-state determination, surface and interface energetics, and AIMD trajectory benchmarking. 
The results show that model performance is property-dependent: for example, \texttt{PET-OMAT-v1 M} gives the lowest relative-energy error, whereas \texttt{UMA OMAT s1p1} provides the best force agreement for the CrCl$_3$/sapphire AIMD trajectory, highlighting the importance of task-specific MLIP selection.

Looking forward, several directions for future development are envisioned. 
First, the integration of molecular dynamics capabilities, including \textit{NVT}, \textit{NPT}, and \textit{NVE} ensembles, would enable the study of finite-temperature properties, diffusion, and phase transitions directly within the platform. 
Second, the addition of NEB calculations for transition-state searches and minimum-energy path determination would broaden the platform's applicability to reaction mechanism studies and catalysis. 
Third, global optimization methods for nanocluster structure prediction represent another natural extension. Fourth, as new MLIP architectures and models continue to emerge at a rapid pace, the modular plugin architecture of MLIP Studio is designed to accommodate their seamless integration with minimal development effort. 
Fifth, the development of automated workflows that chain multiple tasks, for example, geometry optimization followed by vibrational analysis to confirm the nature of a stationary point, or equation of state calculations across multiple models for cross-validation, would further enhance productivity. 
Finally, community-driven contributions of new tasks, post-processing tools, and model integrations are actively encouraged through the official repository~\cite{mlip_studio_github_2025}.

By lowering the barriers to deploying, comparing, and benchmarking foundation MLIPs, MLIP Studio aims to accelerate the adoption of AI and ML methods in everyday computational chemistry and materials science workflows, and to serve as a shared infrastructure for reproducible and transparent evaluation of the rapidly growing ecosystem of foundation models for atomistic simulation.

\color{black}


\section*{Data Availability Statement}
All data generated or analysed during this study, including the input
structures and the numerical data underlying all results, figures, and
tables, are provided in the Supporting Information accompanying this article.
The source code for MLIP Studio is publicly available at
\url{https://github.com/mlipstudio/MLIP-Studio}.

\begin{acknowledgement}

The authors gratefully acknowledge financial support from the Anusandhan National Research Foundation via grant SPR/2023/000397. 
The authors acknowledge the Supercomputer Education and Research Centre (SERC) at the Indian Institute of Science for computational facilities.

Use of Large Language Models (LLMs): OpenAI’s ChatGPT (free version) and Anthropic’s Claude (free version) were used for finding synonyms, exploring alternative framing of sentences, and grammatical refinement. LLMs were not used for factual information or
analysis/interpretation of any result.
Any content generated by LLMs was reviewed and edited carefully.

\end{acknowledgement}
\begin{suppinfo}
Structure files and the output data for the various calculations performed in the article.

\end{suppinfo}

\section*{Conflict of Interest Statement}
The authors declare no conflict of interest.
\bibliography{literature}

\end{document}